\PassOptionsToPackage{hidelinks}{hyperref}
\documentclass[sigconf]{acmart}

\usepackage{array}
\usepackage{calc}
\usepackage{seqsplit}
\usepackage{stfloats} 
\usepackage{placeins} 
\usepackage{float} 
\usepackage{url}
\usepackage{subcaption} 

\providecommand{\tightlist}{%
  \setlength{\itemsep}{0pt}\setlength{\parskip}{0pt}}


\title{The Quality of Claude AI-authored Python Tests Is Not Weaker Than Human-authored Tests}

\author{D. J. Leith}
\affiliation{%
  \institution{Trinity College Dublin}
  \city{Dublin}
  \country{Ireland}
}
\email{doug.leith@tcd.ie}

\acmConference[Preprint]{}{}{}
\acmBooktitle{Preprint}
\acmPrice{}
\acmISBN{}
\acmDOI{}
\renewcommand\footnotetextcopyrightpermission[1]{}



\begin{document}

\begin{abstract}
We evaluate the quality of Claude AI-written Python tests against human-written Python tests from two established open-source projects Django and Pandas. Hundreds of tests per corpus are scored under one identical protocol. Using one-sided non-inferiority bounds, we find that the tests written by recent Claude models (Sonnet/Opus 4.6 and later) are no weaker than the two human-written corpora.  In this study: (i) the AI-written corpus is tests from real tools, not synthetic tests generated in isolation against a fixed target, the setup used by every other AI-test-generation study we are aware of; (ii) every test is individually scored under three independent fault-injection protocols plus a seven-axis qualitative design rubric, allowing methods to cross-validate each other; (iii) tests are scored individually, rather than suite-level, identifying exactly which specific tests need attention.\end{abstract}

\maketitle

\section{Introduction}\label{introduction}

We evaluate the performance of AI-written and human-written Python tests. We find that AI-written tests are no weaker than human-written ones: they are comparable across four independent evaluation methods applied to the same test population. This is contrary to the more pessimistic picture from prior AI-test-generation studies, which report low compile/pass rates and weak oracle accuracy for LLM-generated tests. Note, however, that coding-focused models underwent a well-documented improvement in ability in late 2025\footnote{In November 2025 Claude Opus 4.5 was the first model to break 80\% on SWE-bench Verified.}. The models evaluated here post-date that change, whereas previous work mostly evaluated substantially earlier models.

Testing is what makes AI-written code safe to build on. Large language models write code quickly,
but make mistakes often enough that their output cannot be trusted at face value
\cite{chen2021codex, liu2023evalplus}. The safeguard is
not flawless code but a test suite that locks in intended behaviour, so that a regression
introduced by a later change, AI-made or human-made, is caught rather than silently shipped.
That safeguard is only as good as the tests themselves. Whether AI-written tests are, in practice,
as effective as human-written ones at this job therefore has direct bearing on how safely
AI-assisted development can scale. It is not only an academic interest in code style.

This report scores AI- and human-authored tests under four independent, complementary methods: (i) historical-revert testing, which reverts each test's own real, once-shipped bugfix commit and checks whether the test still fails; (ii) per-commit AST mutation testing, which mutates the lines that same introducing commit itself changed and checks whether the test catches the resulting synthetic fault; (iii) coverage-guided
mutation testing, which places a synthetic fault anywhere in the production code the
test actually executes, without needing a paired commit; and (iv) a seven-axis qualitative rubric that
scores test design directly, independent of whether a particular fault happens to be caught.
This was applied at scale, across five Python corpora: several thousand synthetic mutants and
hundreds of individually scored tests per corpus. The four methods cross-check each other's blind
spots.

To the best of our knowledge, no prior study has evaluated both AI- and human-written tests under a matched, individual-test protocol allowing a direct head-to-head per-test comparison as we do here. This individual-test focus matters beyond precision: it is directly actionable. A suite-level score says a suite is, say, 90\% effective, but not which tests make up the weak 10\%. Scoring each test individually identifies exactly which ones need attention, so effort can be targeted at fixing those tests rather than diffused across a suite that is mostly already sound.

Our main AI-written corpus is also unlike the setup used in every AI-test-generation study we are aware
of. Rather than an LLM generating a test in isolation against a fixed,
already-written function, it is the real test suite of Extractor, a tool built
from scratch by Claude Sonnet 4.6 over several weeks of ordinary, unscripted development. The AI
decided for itself, turn by turn, whether a given moment called for a test. No special
test-writing instructions were given. We also extend the analysis to two large-scale open-source packages with Claude Sonnet/Opus 4.6 AI-authored tests, although for these we do not know the test generation workflow used as clearly as we do for Extractor.

\section{Related work}\label{related-work}

\emph{Whole-test-suite mutation adequacy.} Much of the literature focuses on whether a test suite's mutation score
predicts its ability to catch real faults. Probably the earliest work is \cite{just2014mutants}, see also \cite{petrovic2021mutation}, \cite{papadakis2019mutation}. However,
\cite{papadakis2018correlated} finds the correlation weakens sharply once suite size is controlled for;
\cite{andrews2005mutation} show hand-seeded faults behave differently from real faults;
and \cite{jiaharman2009hom}, \cite{shamshiri2015doautomatically} show
combining mutations does not reliably help, and that generated suites found only
55.7\% of real faults, just 19.9\% of individual suites finding any fault at all i.e.~high scores on
generation-time adequacy translate only weakly into real fault-finding.

\emph{Evaluating the quality of individual tests.} A smaller strand asks whether an individual test
does its job, independent of whole-suite adequacy. In \cite{zhangmesbah2015assertions} it is shown that
assertion quantity and quality, not coverage, predicts effectiveness;
\cite{vahabzadeh2015bugs} finds two-thirds of silently-passing test-code bugs are
assertion-related. The closest precedent for scoring a test against a mutant of its inferred target
is \cite{vercammen2018focal}: on Apache Ant, the whole
suite killed 44 of 55 focal mutants, but the responsible test alone killed only 35.

\emph{Qualitative scoring (``test smells'').} A more mature strand evaluates
test quality by category and severity rather than fault-catching. Its vocabulary originates with \cite{vandeursen2001refactoring}'s ``test smells''. In \cite{bavota2015harmful} it is found that 86\% of JUnit tests carry at least one smell, with
comprehension 30\% worse in its presence; see also survey \cite{garousikucuk2018survey}. Detection is mostly
automated via static rule-matching, but Panichella et al.
\cite{panichella2022smells20years} show developers find such warnings unrepresentative, and
respond better to hand-annotated verified judgement;
\cite{campos2021severity} likewise finds smell severity is a matter of judgement.

\emph{Evaluating AI-written tests.} A fourth, recent strand asks whether LLM-generated tests are as
effective as human-written ones, mostly via compile/pass rates rather than mutation testing.
In \cite{schafer2023empirical} only 48.0\% of
LLM-generated JavaScript tests are found to pass at all; \cite{yuan2024chattester} finds only 24.8\% of
GPT-3.5-generated tests compile and pass before refinement. In
\cite{konstantinou2024oracles} it is found that LLM-written oracles frequently capture actual rather than expected
behaviour, under 50\% oracle accuracy;
\cite{evolution2026residual} finds 99\% of tests that fail after a genuine behavioural change still
pass on the unmodified program (but its Project CodeNet corpus has no AI training-data contamination check). A 2026 head-to-head on real Python
bugs \cite{llmvshuman2026} finds an LLM given the bug report plus fix detects 69\% of faults against 17.2\%
for pre-existing human suites. Patch-review
studies such as METR's \cite{metr2026prs} grade whether AI-generated patches survive maintainer review,
not whether their tests are effective.

\section{Preliminaries}\label{preliminaries}

\subsection{Dataset}\label{dataset}

Data, plus scripts to reproduce the tables and figures here, is available at \url{https://github.com/doug-leith/claude-python-test-quality}

\subsection{Code Corpora Used}\label{code-corpora-used}

We primarily evaluate the test quality in three code corpora: Extractor (apple\_protobuf\_extractor), Django and Pandas; see Table \ref{tab:scale}.

\footnotesize

\begin{table*}[tb]
\footnotesize
\centering
\caption[Corpus scale: authorship, domain, and code/test size for the three corpora studied.]{Corpus scale: authorship, domain, and code/test size\footnotemark{} for the three corpora studied.}\label{tab:scale}
\begin{tabular}{@{}
  >{\raggedright\arraybackslash}p{(\linewidth - 6\tabcolsep) * \real{0.1472}}
  >{\raggedright\arraybackslash}p{(\linewidth - 6\tabcolsep) * \real{0.3672}}
  >{\raggedright\arraybackslash}p{(\linewidth - 6\tabcolsep) * \real{0.2353}}
  >{\raggedright\arraybackslash}p{(\linewidth - 6\tabcolsep) * \real{0.2502}}
@{}}
\toprule
\begin{minipage}[b]{\linewidth}\raggedright
Corpus
\end{minipage} & \begin{minipage}[b]{\linewidth}\raggedright
Written by
\end{minipage} & \begin{minipage}[b]{\linewidth}\raggedright
Domain
\end{minipage} & \begin{minipage}[b]{\linewidth}\raggedright
Size (src/test LOC; tests)
\end{minipage} \\
\midrule
Extractor & Claude (Sonnet 4.6) & ARM64/ObjC reverse eng. & 21k/20.7k; 1,423 tests \\
Django & Human; \textasciitilde88k GitHub stars; active since 2012 & Web framework & 165k/354k; 17,519 tests \\
Pandas & Human; \textasciitilde49k GitHub stars; active since 2010 & Data analysis library & 263k/407k; 20,244 tests \\
\bottomrule
\end{tabular}
\normalsize
\end{table*}
\footnotetext{Note that these LOC numbers include comments/docstrings and white space.}

\normalsize

\emph{Django and Pandas.} Both are large, mature, widely used open-source Python projects with
primarily human-written test suites and long, real commit histories\footnote{See \url{https://github.com/django/django}, \url{https://github.com/pandas-dev/pandas}.}.

\emph{Extractor.} This is Claude Sonnet AI-written code, wholly written by AI from scratch over several weeks of work. It is a real tool (for reverse engineering protobuf descriptors from compiled Apple Objective-C/C binaries), not synthetic test or ``toy'\,' code. Only 22\% of episodes where tests are written/edited trace to a direct ask for
tests; the remaining 78\% were the agent's own initiative e.g.~a new feature or function was requested with no
mention of tests anywhere in the instruction, and the agent wrote tests anyway while delivering the work. In the 22\% of episodes with human prompting, the prompts are always general in nature e.g.~``yes, go ahead with that plan. add tests to ensure good code coverage as you go along.'', ``check the test coverage of the new code, add extra tests if needed.''. See Additional Material for more details.

This contrasts with nearly all of the existing AI-written test quality literature (\S\ref{related-work}), where the AI is handed a fixed, already-written piece of production code (e.g.~a function signature and body, a ``focal method'' plus its class, a method plus a pre-existing test prefix, a self-contained Project CodeNet snippet, or a target function plus a bug patch), and asked only to produce a test for it. The model never writes or evolves the production code itself, and never decides on its own initiative whether a test is warranted. Each test is generated as one isolated, bounded task, not as one moment inside an open-ended, multi-week development process in which the same agent also designs and writes the code under test.

In \S\ref{sec:extras-corpora} we extend the evaluation of AI-authored test quality to Quay and AWX, both large-scale open-source codebases. However, we keep that analysis separate from the Extractor analysis because: (i) we do not know the provenance of the Quay and AWX AI-authored tests as clearly as for Extractor (in particular, we do not know the AI instructions used to generate the tests nor whether, or how much, the tests underwent human revision before being committed), and (ii) the Quay and AWX codebases contain a mix of human and AI authored code/tests whereas Extractor is wholly AI-authored.

\subsection{Test Sampling}\label{test-sampling}

We evaluate tests associated with a production-code change. Tests were drawn uniformly at random from the eligible population, subject to the following exclusion criteria: (i) the evaluation methodology cannot validly score this class of test\footnote{For example, a refactor test whose purpose is to prove old and new implementations are equivalent. Such a test succeeds when it
  passes on both the pre- and post-change code, whereas our test methodologies focus on tests whose correct
  behaviour is to fail when a code regression occurs.}, (ii) the test's failure (or non-run) is an artefact of the test harness or environment, not a real signal\footnote{The test harness is: (a) missing a dependency, (b) has a version mismatch or (c) a build/environment
  limitation that cannot be readily fixed. For example, a pre-2020 \texttt{\seqsplit{asgiref}} default that is a no-op against the modern dependency actually installed, a MariaDB-version-specific check, a compiler-version-only historical fix.}.

\section{Mutation-testing results}\label{mutation-testing-results}

Our goal is to evaluate the quality of AI-authored tests, using human-written tests (Django, Pandas) as a baseline for comparison. The same evaluation protocol is applied to all three corpora i.e.~tests are scored the same way, regardless of which corpus a test comes from. The resulting test scores are therefore directly comparable.

In this section test quality is evaluated under three mutation-based protocols (we consider qualitative evaluation in \S\ref{sec:qualitative}): (i) \emph{historical-revert}: revert the production code change in each test's introducing commit and check whether the test fails, evaluating ``does the test catch the regression'';
(ii) \emph{commit-based mutation}: for the same test/commit pairs, mutate
the code the commit changes and score whether the test catches (``kills'\,') each synthetic mutant;
(iii) \emph{coverage-based mutation}: mutate every production line the test's own execution
is estimated to cover and score whether the test catches each mutant.

Each protocol trades realism for reach in different ways. Historical-revert measures regression against a real, once-shipped defect but only for fixes paired with a test and only as a single pass/fail data point per test. Commit-based mutation keeps the same commit-scoped population but replaces the one real defect with many synthetic ones, trading some realism for a denser, per-test kill-rate signal. Coverage-based mutation testing drops the commit requirement and mutates code the test is estimated to cover, not just a diff. However, that broader reach also means it can mutate code unrelated to the test's actual purpose, so a low kill rate can reflect either weak assertions or simply incidental coupling to code the test was never responsible for. Interpretability also differs sharply. A historical-revert outcome is easy to read: the test either caught a real, once-shipped bug or it did not. A kill-rate value is not so easy to read: there is no established mapping, or even a clear correlation, between a given kill-rate value and real fault tolerance (see \S\ref{sec:interpret-kill-rate}\} below) although, all else being equal, a higher kill rate suggests better fault tolerance than a lower one.

\subsection{Historical-revert testing of individual tests}\label{sec:historical-revert}

\subsubsection{Methodology}\label{methodology-historical-revert}

For each test \texttt{\seqsplit{t}} with introducing (post-fix)
commit \texttt{\seqsplit{c}} touching production files \texttt{\seqsplit{F}}, the harness checks out \texttt{\seqsplit{c}}, confirms \texttt{\seqsplit{t}} currently
passes (baseline), reverts every file in \texttt{\seqsplit{F}} to its pre-fix content from \texttt{\seqsplit{c}}'s parent, and re-runs
\texttt{\seqsplit{t}}. If \texttt{\seqsplit{t}} now fails, the verdict is EFFECTIVE: the test genuinely caught the real historical
regression. If it still passes, the verdict is SURVIVED. A third possible verdict,
UNREACHED, applies when the test structurally can never reach the changed code under any
input its own design could produce at all. The full pseudo-code is in the Additional Material.

\subsubsection{Results}\label{results}

Table \ref{tab:historical-revert-results} shows that the EFFECTIVE rate is consistently high for all three packages. Pandas scores slightly lower than Extractor and Django, and the confidence bounds for all three overlap substantially. The overlap of the confidence intervals means that the data cannot reject the hypothesis that there is ``no difference'' between the corpora. We can also use a one-sided 95\% non-inferiority bound on the difference (Extractor's EFFECTIVE rate - Django/Pandas EFFECTIVE rate) to evaluate the hypothesis that Extractor's EFFECTIVE rate is worse than Django/Pandas. Calculating this\footnote{Bootstrap-resampled, 5,000 resamples with replacement at each corpus's own sample size.} rules out Extractor's EFFECTIVE rate being worse than Django's by more than 3.3\%, and worse than Pandas's by more than 1.4\%, at 95\% confidence.

\footnotesize

\begin{table}[tb]
\footnotesize
\centering
\caption{Historical-revert testing results by corpus.}\label{tab:historical-revert-results}
\begin{tabular}{@{}
  >{\raggedright\arraybackslash}p{(\linewidth - 6\tabcolsep) * \real{0.2802}}
  >{\raggedleft\arraybackslash}p{(\linewidth - 6\tabcolsep) * \real{0.2399}}
  >{\raggedleft\arraybackslash}p{(\linewidth - 6\tabcolsep) * \real{0.2399}}
  >{\raggedleft\arraybackslash}p{(\linewidth - 6\tabcolsep) * \real{0.2399}}
@{}}
\toprule
Metric & Extractor & Django & Pandas \\
\midrule
Scored & 298 & 350 & 350 \\
EFFECTIVE & 288 & 342 & 334 \\
SURVIVED & 9 & 7 & 11 \\
UNREACHED & 1 & 1 & 5 \\
EFFECTIVE Rate & \textbf{96.6\%} & \textbf{97.7\%} & \textbf{95.4\%} \\
95\% CI (Wilson) & 93.9--98.2\% & 95.6--98.8\% & 92.7--97.2\% \\
\bottomrule
\end{tabular}
\normalsize
\end{table}

\normalsize

Two mechanisms account for essentially all SURVIVED and UNREACHED tests across all three corpora:

\begin{enumerate}
\def\labelenumi{\arabic{enumi}.}
\item
  \emph{Weak-value/loose-bound oracle.} The production code path changes when the commit is reverted and the reverted code is executed by the test, but the test's assertion does not distinguish the old behaviour from the new. This is the dominant failure mode in every corpus.
\item
  \emph{Wrong entry point / bypassed wrapper.} The commit changes only one route in the affected logic, and the test exercises a different route. Depending on how much of the changed code sits behind that route, this can be scored as SURVIVED (some of the change is still reached, but not the part that matters) or as UNREACHED (none of it is).
\end{enumerate}

An example from Extractor is a concrete illustration of the first mechanism. The function under test by \texttt{\seqsplit{test\_flags\_struct\_behind\_name\_still\_detected}} decides whether an Objective-C ivar is a ``flags struct'' (a nested struct whose fields are all single-bit flags) by inspecting its raw type-encoding string. Before the commit, this decision used a broad rule. The commit replaces this with a more precise one: a specific marker token must appear, and it may now appear anywhere in the string, not only at the very start. The test's chosen fixture happens to satisfy both rules at once (it starts with the marker token), so reverting the commit does not change the test's result.

\subsection{Commit-based mutation}\label{commit-based-mutation}

\subsubsection{Methodology}\label{methodology}

The population is the same set of commits scored by the historical-revert evaluation above.
Mutations are generated using cosmic-ray v8.4.6\footnote{PIT operates on JVM bytecode. Cosmic-ray and mutmut are the two most actively maintained tools for mutating Python code; other options (e.g.~MutPy, Mutatest) exist but are narrower in scope or largely unmaintained \cite{diallo2024pythontools}. We used cosmic-ray rather than mutmut because it allows control over the mutation operators used.}. Following Laurent et al.'s empirical analysis of PIT's own default operator set \cite{laurent2017pit}, we base mutation on PIT's curated operator set\footnote{This set includes MATH (arithmetic-operator replacement), CONDITIONALS\_BOUNDARY (e.g.~\texttt{\textless{}},\texttt{\textless{}=}), NEGATE\_CONDITIONALS
  (\texttt{\seqsplit{==}},\texttt{\seqsplit{!=}} and similar), INVERT\_NEGATIVES, INCREMENTS (\texttt{\seqsplit{++}},\texttt{\seqsplit{-\/-}}), and
  VOID\_METHOD\_CALL/RETURN\_VALS (deletes/replaces return value).
}.\\
The kill rate reports the percentage of mutants that cause the test to fail i.e.~that are detected by the test and so ``killed''.

\subsubsection{Results}\label{results-1}

Table \ref{tab:mutation-results-combined} shows that the confidence bounds of the kill rate overlap for all three corpora. The one-sided 95\% non-inferiority bound calculated by bootstrapping rules out Extractor's kill rate being worse than Django's by more than 0.5\%, at 95\% confidence. Against Pandas the bound is positive (+2.4\%), which rules out Extractor's kill rate being worse at all i.e.~the data is 95\% confident Extractor's kill rate is at least 2.4\% higher than Pandas. (The table also shows the coverage-based mutation results discussed below, \S\ref{coverage-based-mutation}, for direct comparison.)

\footnotesize

\begin{table}[tb]
\footnotesize
\centering
\caption[AST mutation testing results by corpus, per-commit and coverage-based protocols\textsuperscript{,}.]{AST mutation testing results by corpus, per-commit and coverage-based protocols\footnotemark{}\textsuperscript{,}\footnotemark{}.}\label{tab:mutation-results-combined}
\begin{tabular}{@{}
  >{\raggedright\arraybackslash}p{(\linewidth - 6\tabcolsep) * \real{0.2802}}
  >{\raggedleft\arraybackslash}p{(\linewidth - 6\tabcolsep) * \real{0.2399}}
  >{\raggedleft\arraybackslash}p{(\linewidth - 6\tabcolsep) * \real{0.2399}}
  >{\raggedleft\arraybackslash}p{(\linewidth - 6\tabcolsep) * \real{0.2399}}
@{}}
\toprule
Metric & Extractor & Django & Pandas \\
\midrule
\multicolumn{4}{>{\raggedright\arraybackslash}p{\linewidth-2\tabcolsep}}{\textbf{Per-commit}}\\
Commits & 48 & 100 & 116 \\
Tests & 234 & 100 & 116 \\
Scored mutants & 695 & 273 & 291 \\
Killed & 370 & 131 & 131 \\
Survived & 325 & 142 & 160 \\
Kill rate & \textbf{53.2\%} & \textbf{48.0\%} & \textbf{45.0\%} \\
95\% CI (Wilson) & 49.5--56.9\% & 42.1--53.9\% & 39.4--50.8\% \\
\midrule
\multicolumn{4}{>{\raggedright\arraybackslash}p{\linewidth-2\tabcolsep}}{\textbf{Coverage-based}}\\
Tests & 249 & 342 & 344 \\
Scored mutants & 2,791 & 2,744 & 1,616 \\
Killed & 1,293 & 1,373 & 738 \\
Survived & 1,498 & 1,371 & 878 \\
Kill rate & \textbf{46.3\%} & \textbf{50.0\%} & \textbf{45.7\%} \\
95\% CI (Wilson) & 44.5--48.2\% & 48.2--51.9\% & 43.3--48.1\% \\
\bottomrule
\end{tabular}
\normalsize
\end{table}
\footnotetext{Observe that for Extractor the number of tests is larger than the number of commits. Django and Pandas had one test per commit, whereas Extractor often included multiple tests in one commit.}
\footnotetext{The number of tests scored is different between the per-commit and coverage-based evaluations. This is because production code with no operators to mutate is not included in the counts. The per-commit protocol mutates the commit code while the coverage-based protocol mutates code estimated from test coverage. Often the code estimated from test coverage was found to contain more mutable operators than the commit code, and so the number of tests evaluated is higher.}

\normalsize

\subsubsection{Interpreting kill rate}\label{sec:interpret-kill-rate}

Despite its popularity, mutation kill rate has no validated absolute scale. The intuition behind treating ``higher is better'' as a universal rule is that a suite killing more mutants must be more thorough. Thoroughness, however, in the sense that matters (catching real bugs), is exactly the property the literature finds weakest. Papadakis et al. \cite{papadakis2018correlated} show that once test suites are compared at matched sizes, the correlation between mutation score and real fault detection is weak. Instead they show that when test suites are ranked by kill rate, higher-ranked suites tend to have better fault detection than the full population of same-size suites. That is, the kill rate's ordinal information (higher kill rate suites tend to be better, relative to peers of the same size) remains useful. With this in mind, the results here suggest that the Extractor, Django and Pandas test suites can be expected to provide similar levels of fault protection.

\subsection{Coverage-based mutation}\label{coverage-based-mutation}

\subsubsection{Methodology}\label{methodology-1}

For each test the production code executed by the test is estimated using a custom coverage analysis tool. That code is then mutated using cosmic-ray and the kill-rate reported.
Coverage-based mutation raises a scoping question the per-commit protocol avoids: a test's own call tree can pass through production code that is just scaffolding (e.g.~fixtures, builders, mocks, setup/teardown) and has nothing to do with what the test is actually checking. Our coverage tool tries to detect and exclude such scaffolding. It uses two strategies for this: (i) exclude code outside the module(s) a test directly imports, (ii) try to identify shared frameworks/libraries that most tests in the corpus execute incidentally and exclude these.

\subsubsection{Results}\label{results-2}

Table \ref{tab:mutation-results-combined} shows that the confidence bounds of the kill rate overlap for all three corpora. The one-sided 95\% non-inferiority bound calculated by bootstrapping rules out Extractor's kill rate being worse than Django's by more than 5.9\%, at 95\% confidence, and rules out being more than 2\% worse than Pandas. Bearing in mind the discussion in \S\ref{sec:interpret-kill-rate}, the data suggests that the test suites of all three corpora provide similar levels of fault protection.

\subsection{Whole package mutation}\label{whole-package-mutation}

Our main interest here is in evaluating the quality of individual tests. However, in the literature it is more common to instead apply mutations to the whole code-base of a package and evaluate the ability of the whole test-suite to detect these mutations. For completeness, Table \ref{tab:whole-suite-results} shows the measured full code-base/whole test-suite mutation kill rates for Extractor and the django/utils/ component of Django (chosen because it is roughly the same size as the Extractor code). It can be seen that the Extractor and Django test-suites have almost identical kill rates.

\footnotesize

\begin{table}[tb]
\footnotesize
\centering
\caption{Whole-test-suite mutation kill rate, Extractor vs.~Django utils package.}\label{tab:whole-suite-results}
\begin{tabular}{@{}
  >{\raggedright\arraybackslash}p{(\linewidth - 4\tabcolsep) * \real{0.3686}}
  >{\raggedleft\arraybackslash}p{(\linewidth - 4\tabcolsep) * \real{0.3157}}
  >{\raggedleft\arraybackslash}p{(\linewidth - 4\tabcolsep) * \real{0.3157}}
@{}}
\toprule
Metric & Extractor & Django \\
\midrule
Scored & 679 & 679 \\
Killed & 399 & 396 \\
Survived & 280 & 283 \\
Kill rate & \textbf{58.8\%} & \textbf{58.3\%} \\
95\% CI (Wilson) & 55.0--62.4\% & 54.6--62.0\% \\
\bottomrule
\end{tabular}
\normalsize
\end{table}

\normalsize

\section{Qualitative test-quality results}\label{sec:qualitative}

The mutation-testing analysis above tries to answer ``does the test catch a wrong implementation'', a
behavioural, pass/fail question. In this section we consider a separate, complementary qualitative evaluation that asks design-quality questions e.g.~is a test as precise as it could be without becoming fragile, does it couple to implementation incidentally, does it exhibit a known anti-pattern, does it check one cohesive claim.

\footnotesize

\begin{table*}[tb]
\footnotesize
\centering
\caption{The seven-axis qualitative rubric: what each axis asks, and its verdict scale.}\label{tab:rubric-axes}
\begin{tabular}{@{}
  >{\raggedright\arraybackslash}p{(\linewidth - 4\tabcolsep) * \real{0.2001}}
  >{\raggedright\arraybackslash}p{(\linewidth - 4\tabcolsep) * \real{0.3249}}
  >{\raggedright\arraybackslash}p{(\linewidth - 4\tabcolsep) * \real{0.4750}}
@{}}
\toprule
\begin{minipage}[b]{\linewidth}\raggedright
Axis
\end{minipage} & \begin{minipage}[b]{\linewidth}\raggedright
Verdict scale
\end{minipage} & \begin{minipage}[b]{\linewidth}\raggedright
What it asks
\end{minipage} \\
\midrule
Intent fidelity & \texttt{\seqsplit{MATCH}} / \texttt{\seqsplit{PARTIAL}} / \texttt{\seqsplit{MISMATCH}} & Does the assertion actually check what the docstring/name/comment claims? \\
Assertion strength & \texttt{\seqsplit{STRICT}} / \texttt{\seqsplit{MODERATE}} / \texttt{\seqsplit{WEAK}} / \texttt{\seqsplit{n/a}} & Is the assertion at the ceiling of precision the claim calls for, without introducing fragility it does not need? \\
Behaviour/implementation coupling & \texttt{\seqsplit{BEHAVIOR}} / \texttt{\seqsplit{JUSTIFIED-MECHANISM}} / \texttt{\seqsplit{UNJUSTIFIED-COUPLING}} & Does it assert an externally observable outcome, or an incidental implementation detail that would break under a legitimate refactor? \\
Anti-pattern & present / absent per category & A checklist for named smells: dead lever, assertion masking (coverage-real or diagnostic-only), over-broad mock, swallowed exception, log-instead-of-assert, tautology, skip/xfail masking \\
Setup/fixture proportionality & \texttt{\seqsplit{OK}} / \texttt{\seqsplit{DEFECTIVE}} & Does Arrange build only what is needed, and does it construct the scenario the test actually claims to exercise? \\
Purpose stated & \texttt{\seqsplit{CLEAR}} (per-test/class/module/external) / \texttt{\seqsplit{PARTIAL}} / \texttt{\seqsplit{WRONG}} / \texttt{\seqsplit{NONE}} & Does some source (docstring, comment, commit message, or a linked issue tracker) state what the test verifies and why? \\
Cohesion & \texttt{\seqsplit{SINGLE}} / \texttt{\seqsplit{BUNDLED}} & Does the test check one cohesive claim, or bundle two or more independent claims into one method? \\
\bottomrule
\end{tabular}
\normalsize
\end{table*}

\normalsize

Scoring uses a seven-axis rubric, shown in Table \ref{tab:rubric-axes}, and is carried out by AI agents.
The instructions given to the agents, including the full definition of each
category above, are given in the Additional Material. The defects in all tests categorised as having critical flaws are verified by human manual analysis.

Table \ref{tab:severity-breakdown-per-corpus} summarises the qualitative evaluation results, mapping the non-ideal scoring verdicts to four severity tiers: critical / moderate / minor / informational. It can be seen that the coupling, purpose-stated and setup/fixture axes generally raise few issues. Anti-patterns are relatively common in all three corpora. Cohesion issues are infrequent in the Extractor tests, but more common in Django and especially Pandas. In all three corpora tests are flagged as having intent-fidelity and assertion-strength issues, and these tend to be more common for Extractor.

A bootstrapped one-sided non-inferiority bound rules out Extractor's critical rate exceeding Django's by more than 2.3\% and Pandas's by more than 3.2\%. More importantly,
the absolute number of tests flagged is small. Overall only 12 tests are flagged as critical (6/5/1 for Extractor/Django/Pandas). These counts are small enough that the cross-corpus comparison should be read cautiously: a handful of borderline moderate/critical calls going the other way would change the ordering. Note also that there are likely to be false negatives (i.e.~test defects that are not discovered), see \S \ref{sec:test-triage}.

\footnotesize

\begin{table}[tb]
\scriptsize
\centering
\caption[Per-axis severity breakdown, by corpus.]{Per-axis severity breakdown, by corpus\footnotemark{}.}\label{tab:severity-breakdown-per-corpus}
\begin{tabular}{@{}
  >{\raggedright\arraybackslash}p{(\linewidth - 14\tabcolsep) * \real{0.1481}}
  >{\raggedleft\arraybackslash}p{(\linewidth - 14\tabcolsep) * \real{0.1391}}
  >{\raggedleft\arraybackslash}p{(\linewidth - 14\tabcolsep) * \real{0.1524}}
  >{\raggedleft\arraybackslash}p{(\linewidth - 14\tabcolsep) * \real{0.1016}}
  >{\raggedleft\arraybackslash}p{(\linewidth - 14\tabcolsep) * \real{0.1245}}
  >{\raggedleft\arraybackslash}p{(\linewidth - 14\tabcolsep) * \real{0.0982}}
  >{\raggedleft\arraybackslash}p{(\linewidth - 14\tabcolsep) * \real{0.1344}}
  >{\raggedleft\arraybackslash}p{(\linewidth - 14\tabcolsep) * \real{0.1016}}
@{}}
\toprule
\begin{minipage}[b]{\linewidth}\raggedright
\end{minipage} & \begin{minipage}[b]{\linewidth}\raggedleft
Intent-fidelity
\end{minipage} & \begin{minipage}[b]{\linewidth}\raggedleft
Assertion strength
\end{minipage} & \begin{minipage}[b]{\linewidth}\raggedleft
Coupling
\end{minipage} & \begin{minipage}[b]{\linewidth}\raggedleft
Anti-pattern
\end{minipage} & \begin{minipage}[b]{\linewidth}\raggedleft
Setup
\end{minipage} & \begin{minipage}[b]{\linewidth}\raggedleft
Purpose-stated
\end{minipage} & \begin{minipage}[b]{\linewidth}\raggedleft
Cohesion
\end{minipage} \\
\midrule
\multicolumn{8}{>{\raggedright\arraybackslash}p{\linewidth-2\tabcolsep}}{\textbf{Extractor (n=298, critical=6 (2.0\%), moderate=58 (19.4\%))}}\\
Critical & 4 & 6 & 0 & 1 & 2 & 0 & 0 \\
Moderate & 9 & 37 & 0 & 18 & 1 & 1 & 1 \\
Minor & 0 & 0 & 0 & 5 & 0 & 3 & 0 \\
Info & 0 & 0 & 11 & 0 & 0 & 5 & 0 \\
Total & 13 & 43 & 11 & 24 & 3 & 9 & 1 \\
\midrule
\multicolumn{8}{>{\raggedright\arraybackslash}p{\linewidth-2\tabcolsep}}{\textbf{Django (n=342, critical=5 (1.5\%), moderate=44 (12.9\%))}}\\
Critical & 2 & 2 & 0 & 0 & 2 & 0 & 0 \\
Moderate & 5 & 36 & 0 & 3 & 1 & 0 & 8 \\
Minor & 0 & 0 & 0 & 4 & 0 & 79 & 0 \\
Info & 0 & 0 & 28 & 0 & 0 & 149 & 0 \\
Critical+\allowbreak Moderate & 7 & 38 & 0 & 3 & 3 & 0 & 8 \\
\midrule
\multicolumn{8}{>{\raggedright\arraybackslash}p{\linewidth-2\tabcolsep}}{\textbf{Pandas (n=344, critical=1 (0.3\%), moderate=51 (14.8\%))}}\\
Critical & 0 & 1 & 0 & 0 & 0 & 0 & 0 \\
Moderate & 6 & 29 & 2 & 1 & 5 & 13 & 21 \\
Minor & 0 & 0 & 0 & 1 & 0 & 5 & 0 \\
Info & 0 & 0 & 18 & 0 & 0 & 56 & 0 \\
Total & 6 & 30 & 20 & 2 & 5 & 74 & 21 \\
\bottomrule
\end{tabular}
\normalsize
\end{table}
\footnotetext{Note that a test is counted under every axis where it scored non-ideal, so two independently real properties of the same test, e.g.~a weak assertion and a thin docstring, are counted separately in the row tallies. The headline numbers, e.g.~critical=6 for Extractor, report the actual number of tests affected.}

\normalsize

\subsection{Curated exemplars}\label{curated-exemplars}

This section presents a handful of illustrative findings, tagged by \texttt{\seqsplit{test\_id}}. These are
not the full data set, they are a ``vivid, compelling'' subset (Braun \& Clarke's Phase 6) chosen to
make each axis's numbers legible, not just countable.

\subsubsection{Detailed example}\label{detailed-example}

Before the per-axis entries below, we present one Extractor finding worked through in full to show the kind
of evidence used to arrive at a \texttt{\seqsplit{critical}} tag. It is not enough that a potential defect is identified, it needs to be confirmed by mutating the code and a fix demonstrated.

\emph{Test}: \texttt{\seqsplit{test\_resolve\_image\_exact\_full\_path}}

{\footnotesize
\begin{verbatim}
def test_resolve_image_exact_full_path(self):
    """An exact full path resolves via tier 1
    even when a shorter name would too."""
    full = '/System/Library/Frameworks/' \
           'Foundation.framework/Foundation'
    path = self._make_cache([(0x181200000, full)])
    with DyldCacheReader(path) as r:
        self.assertEqual(
            r.resolve_image(full), 0x181200000)
\end{verbatim}
}

The function under test, \texttt{\seqsplit{resolve\_image}}, tries three tiers in order and returns at the first to
yield a candidate: (1) exact full-path match, (2) framework-principal-binary identity, (3) unique
substring.

\emph{Claim vs.~check}: the test docstring claims this proves tier 1 wins ``even when a shorter name would
too'', a priority claim. But the fixture builds a cache with exactly one image and queries it
with that image's own full path. Tier 3 (``unique substring'') trivially matches too since a string is
always a substring of itself. With only one image present, whichever tier actually resolves the
query returns the identical address, so the test cannot distinguish ``tier 1 fired'' from ``tier 1 is
broken and tier 3 silently covered for it.''

\emph{Live proof}: tier 1 was commented out entirely in a dedicated pinned worktree at this test's own
introducing commit, and the real test rerun. It still passed: tier 3's substring check returns
the same address the deleted tier 1 would have.

\emph{The fix}: add a second image whose path merely contains the queried string, making tier 3 ambiguous, so only tier 1's exact match can resolve the original query uniquely.

This detail also makes human verification of the AI analysis relatively straightforward.

\subsubsection{Intent-fidelity}\label{intent-fidelity}

\emph{\texttt{\seqsplit{test\_json\_response\_raises\_type\_error\_with\_safe\_arg}}} (Django, critical). The
test's docstring claims two things: bad input raises \texttt{\seqsplit{TypeError}}, and passing
\texttt{\seqsplit{safe=True}} at all raises a deprecation warning. Only the first is actually checked (the
mechanism why is covered under axis 4, below). The second claim goes untested, a mismatch between
what's claimed and what's verified, which is why this is scored critical.

\subsubsection{Assertion strength}\label{assertion-strength}

\emph{\texttt{\seqsplit{test\_resolve\_image\_exact\_full\_path}}} (Extractor, critical), see above.

\subsubsection{Behaviour/implementation coupling}\label{behaviourimplementation-coupling}

\emph{\texttt{\seqsplit{test\_has\_key\_race\_handling}}} (Django, \texttt{\seqsplit{JUSTIFIED-MECHANISM}}). The assertion checks
\texttt{\seqsplit{mocked\_open.assert\_called\_once()}}, an implementation detail that looks at first glance like
exactly the kind of coupling axis 3 penalizes. But the bug this test guards against is
specifically about which low-level mechanism closes a TOCTOU (time-of-check-to-time-of-use)
race: catching \texttt{\seqsplit{open()}}'s own exception, versus checking a file's existence and opening it as two
separate steps, which leaves a window for another process to change the file in between. Here the
mechanism is the behaviour being promised, so testing it directly is the right call, not a
smell.

\subsubsection{Anti-pattern}\label{anti-pattern}

\emph{\texttt{\seqsplit{test\_json\_response\_raises\_type\_error\_with\_safe\_arg}}} (Django, critical). Combines
\texttt{\seqsplit{assertRaisesMessage}}/\texttt{\seqsplit{assertWarnsMessage}} in one \texttt{\seqsplit{with\ (A,\ B):}} tuple, checking for a \texttt{\seqsplit{TypeError}}
and a deprecation warning at the same time:

{\footnotesize
\begin{verbatim}
with (
  self.assertRaisesMessage(TypeError,...),
  self.assertWarnsMessage(RemovedInDjango71Warning,...),
):
  JsonResponse([1, 2, 3], safe=True)
\end{verbatim}
}

Python exits stacked context managers in reverse order: when the code raises, the warning check
exits first, sees an exception already in flight, and defers to it without ever checking whether
its own warning fired. The warning check is present in the source and looks like it should catch
a regression, but structurally can never run its own pass/fail logic -- a so-called \texttt{\seqsplit{dead\ lever}}.
The test would not notice if the deprecation warning it was explicitly written
to guard were deleted entirely.

\subsubsection{Setup/fixture proportionality}\label{setupfixture-proportionality}

\emph{\texttt{\seqsplit{test\_forward\_branch\_handler\_dominance}}} (Extractor, critical).
This test's setup builds two branches of an if/else, but never rejoins them into a single path afterward. The logic under test exists specifically to determine which earlier branch controls a given later point in the code -- and that's only genuinely hard to get right at the point where branches rejoin. Because the setup never creates that rejoined case, nothing the test checks can catch a wrong answer. That's a setup problem, not just a weak assertion: the scenario it builds is incapable of exercising the one case that matters.

\subsubsection{Purpose stated}\label{purpose-stated}

\emph{\texttt{\seqsplit{test\_read\_index\_col\_none}}} (Pandas, minor). The only comment on this test (``GH 7369, make sure
can read a 0-obs dta file'') is copied verbatim from the sibling test \texttt{\seqsplit{test\_read\_empty\_dta}}
immediately above it in the same class. The text describes that other test's scenario, not this
one's: this test's DataFrame has 5 rows, not 0, and its actual claim, confirmed by the
introducing commit PR \#49745, is that \texttt{\seqsplit{index\_col=None}} should produce a \texttt{\seqsplit{RangeIndex}}.
Following the comment's own issue reference makes things worse, not better: GH 7369 is the
empty-dta-file bug, an entirely different issue than the one this test actually guards against.

\subsubsection{Cohesion}\label{cohesion}

\emph{\texttt{\seqsplit{test\_json\_response\_raises\_type\_error\_with\_safe\_arg}}} (Django, critical). Bundles a
validation claim (bad input raises \texttt{\seqsplit{TypeError}}) with an unrelated deprecation claim (using
\texttt{\seqsplit{safe=}} at all warns) into one method, and this is causal, not merely correlational: it is
exactly this bundling that forces the compound \texttt{\seqsplit{with\ (A,\ B):}} whose exit-order semantics produce
the axis-4 dead lever above. Two separate test methods would have made the bug structurally
impossible, independent of getting the context-manager order right.

\section{Test Triage}\label{sec:test-triage}

Scoring individual tests, not just test-suites in aggregate, lets us flag the specific tests worth a closer look and also surface common patterns among them. It also lets us check whether different evaluation methodologies, e.g.~the mutation-based and qualitative methods, agree on which tests are problematic or flag different ones.

\subsection{Comparing historical-revert and qualitative scoring}\label{sec:compare}

Table \ref{tab:triage-overlap} summarises the number of tests flagged as problematic only by the historical-revert evaluation, only by the qualitative evaluation, and flagged by both. Also shown are the number of those tests that manual analysis confirm as having defects\footnote{See \S \ref{sec:test-details} for a full list of tests flagged, together with commentary on the actual behaviour of each test.}.

\begin{table}[tb]
\footnotesize
\centering
\caption{Tests confirmed genuinely defective / tests flagged, by which methodology flagged them, by corpus. ``Qualitative'' = assertion-strength WEAK; ``Historic-revert'' = SURVIVED/UNREACHED (non-EFFECTIVE).}\label{tab:triage-overlap}
\begin{tabular}{@{}
  >{\raggedright\arraybackslash}p{(\linewidth - 6\tabcolsep) * \real{0.1955}}
  >{\raggedleft\arraybackslash}p{(\linewidth - 6\tabcolsep) * \real{0.2524}}
  >{\raggedleft\arraybackslash}p{(\linewidth - 6\tabcolsep) * \real{0.2607}}
  >{\raggedleft\arraybackslash}p{(\linewidth - 6\tabcolsep) * \real{0.2914}}
@{}}
\toprule
Corpus & Flagged by both & Qualitative only & Historic-revert only \\
\midrule
Extractor & 2/2 (100\%) & 5/5 (100\%) & 2/6 (33.3\%) \\
Django & 3/3 (100\%) & 6/6 (100\%) & 2/12 (16.7\%) \\
Pandas & 0/0 (--) & 3/3 (100\%) & 7/15 (46.7\%) \\
\bottomrule
\end{tabular}
\normalsize
\end{table}

Seven Extractor tests are flagged by the qualitative evaluation and all are confirmed to have defects. Of these seven, only two fail as SURVIVED by historical-revert, the rest are flagged by historical-revert as EFFECTIVE\footnote{Three as EFFECTIVE (TESTING\_NEW\_FUNCTION). This occurs for a commit that introduces an entirely new class/module, or a new method/field/parameter on an already-existing class or function, plus associated tests. Reverting the change to the production code leaves the tests with no function/module/method to execute and so they fail to execute. This is categorised as EFFECTIVE because the revert can be trivially detected.}.
Eight Extractor tests are flagged as non-EFFECTIVE by historical-revert but four are genuinely flawed. Two of the flawed tests were flagged by both the historical-revert and qualitative evaluations. The other two flawed tests are not flagged by the qualitative evaluation. The other four tests are correctly found to be well-written by the qualitative evaluation.

\footnotesize

\begin{table}[tb]
\footnotesize
\centering
\caption{False-positive rate of each triage methodology, by corpus.}\label{tab:fp-rate-comparison}
\begin{tabular}{@{}
  >{\raggedright\arraybackslash}p{(\linewidth - 4\tabcolsep) * \real{0.2427}}
  >{\raggedleft\arraybackslash}p{(\linewidth - 4\tabcolsep) * \real{0.3527}}
  >{\raggedleft\arraybackslash}p{(\linewidth - 4\tabcolsep) * \real{0.4046}}
@{}}
\toprule
Corpus & Qualitative FP rate & Historical-revert FP rate \\
\midrule
Extractor & 0/7 (0\%) & 4/8 (50.0\%) \\
Django & 0/9 (0\%) & 10/15 (66.7\%) \\
Pandas & 0/3 (0\%) & 8/15 (53.3\%) \\
\bottomrule
\end{tabular}
\normalsize
\end{table}

\normalsize

Similar behaviour is observed for Django and Pandas. Table \ref{tab:fp-rate-comparison} gives each methodology's false-positive (FP) rate i.e.~of the tests each flags as problematic,
how many turn out, on independent inspection, to have no real defect. The false positive rate is much higher for the historical-revert evaluation. This reflects the fact that a single mutation (reverting the production code changes in the commit containing the test) is often not discriminating enough to accurately evaluate overall test quality. The qualitative evaluation does not produce false positives, but it can produce false negatives: tests which have defects but which are assigned an assertion-strength of STRICT.

\subsection{AST-mutation scoring}\label{ast-mutation-scoring}

\begin{figure}[tb]
\centering
\begin{subfigure}{0.49\linewidth}
\centering
\includegraphics[width=\linewidth,keepaspectratio]{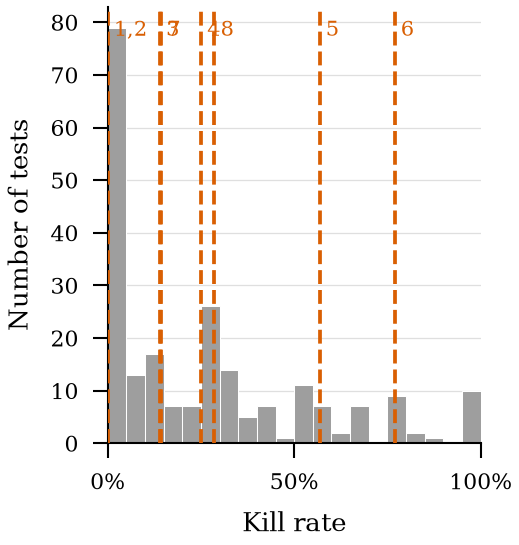}
\caption{Per-commit.}\label{fig:killrate-distributions-a}
\end{subfigure}
\hfill
\begin{subfigure}{0.49\linewidth}
\centering
\includegraphics[width=\linewidth,keepaspectratio]{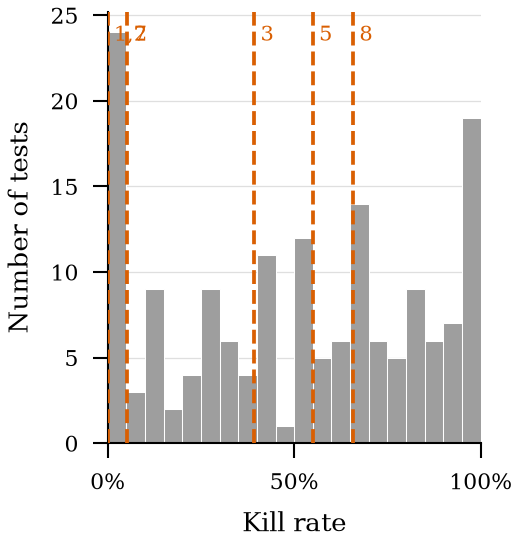}
\caption{Coverage-guided.}\label{fig:killrate-distributions-b}
\end{subfigure}
\caption{Kill-rate distributions for Extractor, with defective tests marked by dashed line.}\label{fig:killrate-distributions}
\end{figure}

Figure \ref{fig:killrate-distributions} shows the distribution of kill rates across all tests and the dashed lines indicate the kill rates of the nine Extractor tests\footnote{The per-commit plot shows 8 tests, the coverage-guided 6 tests. This is because the commit/code-covered by the other tests contained no operators to mutate i.e.~this is correct behaviour.} found to have a real defect in \S \ref{sec:compare}. Neither distribution cleanly separates defective from non-defective tests. It is not clear whether this is because the kill rate is not informative of test defects, or because the test defects uncovered by historical-revert and qualitative evaluation are an incomplete, possibly biased, sample.

As already discussed above, there is good reason to suspect that kill rate is, at best, only weakly correlated with test quality\footnote{We can see how this might be the case by looking at a concrete example: Extractor's \texttt{\seqsplit{test\_word\_substitution}} kills 57\% of its commit's mutations and 55\% of its coverage-guided mutations, neither figure an obvious red flag. The production code this test exercises has two structurally distinct branches, a lowercase and an uppercase branch, but the test is defective because it only checks the uppercase branch. Of the 37 mutants tested, 35 modify code surrounding the branch but not the branch itself. One mutant modifies the uppercase branch and is killed, one modifies the lowercase branch and survives. The raw kill rates therefore tell us little about the defect in the test, only inspecting which mutants survive or reading the code directly can do this.}.

Nevertheless, while a low kill rate might only be a weak signal it might still be useful as a filter to narrow down a set of tests for closer inspection. To investigate this, we took the 25 lowest kill-rate Extractor tests under the per-commit and coverage-guided mutation evaluations (42 unique tests, some appearing in both evaluations) and used AI agents to analyse these for the failure mechanisms above plus any other defect pattern that the survived mutants suggested (see Additional Material for instructions given to the agents). Five tests were found to have a defect, three not flagged by the mutation or qualitative evaluations\footnote{See Additional Material \S\ref{sec:lowkillrate-detail} for details, including manual analysis of the defects.}. That is, of the 42 low kill rate tests examined, 12\% turn out to have defects. For comparison, we also carried out the same analysis of the 25 \emph{highest} kill-rate Extractor tests and found no defective tests. A low kill rate therefore does indeed seem to be a useful filter for test triage. Its usefulness is not confined to Extractor. Table \ref{tab:low-killrate-triage-summary} shows that a similar filter also finds defects by examining the 25 tests with lowest kill rates in Django and Pandas, though at a lower rate for Django. Five of the six defects found this way were independently found by the qualitative evaluation, only one (t\texttt{\seqsplit{est\_groupby\_resample\_kind}} in Pandas) was missed by both the mutation and qualitative evaluations.

All but one of the Django and Pandas tests flagged are new i.e.~not flagged by the historic-commit or qualitative evaluations.

\footnotesize

\begin{table}[tb]
\scriptsize
\centering
\caption{Low-kill-rate triage defect-find rate, by corpus.}\label{tab:low-killrate-triage-summary}
\begin{tabular}{@{}
  >{\raggedright\arraybackslash}p{(\linewidth - 6\tabcolsep) * \real{0.2554}}
  >{\raggedleft\arraybackslash}p{(\linewidth - 6\tabcolsep) * \real{0.2159}}
  >{\raggedleft\arraybackslash}p{(\linewidth - 6\tabcolsep) * \real{0.2644}}
  >{\raggedleft\arraybackslash}p{(\linewidth - 6\tabcolsep) * \real{0.2644}}
@{}}
\toprule
Corpus & Per-commit & Coverage-guided & Combined unique \\
\midrule
Extractor (AI) & 3/25 & 3/25 & 5/42 \\
Pandas (human) & 4/25 & 2/25 & 5/46 \\
Django (human) & 0/25 & 1/25 & 1/50 \\
\bottomrule
\end{tabular}
\normalsize
\end{table}

\normalsize

\section{Results for other AI-authored corpora}\label{sec:extras-corpora}

The results above are for one AI-authored codebase, Extractor. Since we developed Extractor ourselves, its provenance is clear. In particular, we know that the tests were wholly AI written with no human review/\allowbreak modification and no detailed human prompting/\allowbreak instructions.

In this section we extend the analysis to two open-source codebases with AI-authored tests, Quay and AWX\footnote{See \url{https://github.com/quay/quay}, \url{https://github.com/ansible/awx}.}.

\subsection{Search for AI-authored code}\label{search-for-ai-authored-code}

We initially carried out a search of github for open source Python packages that are wholly AI authored, i.e.~similar to Extractor. We cloned candidate repositories for closer inspection of their commit history, code size, language used etc. We also searched chat groups focussed on AI code development\footnote{Hacker News, Reddit r/ClaudeAIm r/LocalLLaMA} for discussion of such code, curated ``awesome-vibe-coding''/``awesome-ralph''/``awesome-claude-code'' lists, as well as used direct web search. We found that repositories that explicitly brand themselves ``vibe coded'' or ``written entirely by Claude'' are overwhelmingly ``toy'' scale i.e.~a few hundred of lines with no real test suite.

We then revised the search to focus on codebases: (i) with more than 100 AI-authored Python tests (other parts of the code, including some of the tests, could be human-authored), (ii) PR's and/or commits with a ``Co-authored-by: Claude'' tag and stating version Opus/Sonnet 4.6 or later, and (iii) code and tests are written together (rather than tests being back-fitted to existing code). In addition to direct search of github we also searched the AIDev dataset (933K agent-authored PRs), but the content of this is too old to cover Opus/Sonnet 4.6. These searches turned up two main candidates: Quay and AWX. Other codebases, including RNAlysis, ha-pitboss, pypureomapi and ouroboros, were found to have fewer than 100 AI-written tests. The Paddleformers\footnote{\url{https://github.com/PaddlePaddle/PaddleFormers}} codebase by Baidu met criteria (i) and (ii) but was excluded because analysis of the commits makes it clear that the tests are back-fitted to existing code. We discuss the restriction to the recent\footnote{Opus 4.6 was released on 5th Feb 2026, Sonnet 4.6 on 17th Feb 2026.} Opus 4.6/Sonnet 4.6 and later models further below.

Quay (quay/quay) is an open-source container image registry supporting Docker Registry and OCI, with authentication, ACLs, geo-replicated storage, and security scanning. It is maintained by Red Hat, which also uses it to run the public quay.io registry service. AWX (ansible/awx) is an open-source web-based UI, REST API, and task engine for running and scheduling Ansible playbooks (Ansible is an open-source IT automation tool that configures systems, deploys software, and orchestrates multi-machine tasks via YAML ``playbooks'') and is one of the upstream projects for Red Hat's Ansible Automation Platform.

In both packages there are a substantial number of recent commits that contain new tests carrying a ``Co-authored-by: Claude'' tag on the commit, see Table \ref{tab:extras-scale}.

\begin{table}[tb]
\footnotesize
\centering
\caption{Details of Quay and AWX.}\label{tab:extras-scale}
\begin{tabular}{@{}
  >{\raggedright\arraybackslash}p{(\linewidth - 6\tabcolsep) * \real{0.1793}}
  >{\raggedright\arraybackslash}p{(\linewidth - 6\tabcolsep) * \real{0.2301}}
  >{\raggedright\arraybackslash}p{(\linewidth - 6\tabcolsep) * \real{0.1945}}
  >{\raggedright\arraybackslash}p{(\linewidth - 6\tabcolsep) * \real{0.3962}}
@{}}
\toprule
\begin{minipage}[b]{\linewidth}\raggedright
Corpus
\end{minipage} & \begin{minipage}[b]{\linewidth}\raggedright
Written by
\end{minipage} & \begin{minipage}[b]{\linewidth}\raggedright
Domain
\end{minipage} & \begin{minipage}[b]{\linewidth}\raggedright
Size (src/test/AI-test LOC; tests/AI-tests
\end{minipage} \\
\midrule
Quay (quay/quay) & Claude Opus 4.6, Sonnet 4.6 & Cloud infrastructure & \ensuremath{\sim}128,300/\allowbreak \ensuremath{\sim}79,700/\allowbreak \ensuremath{\sim}3,600, \ensuremath{\sim}3,800/214 \\
AWX (ansible/awx) & Claude Opus 4.6/4.8, Fable 5 & IT automation & \ensuremath{\sim}104,900/\allowbreak \ensuremath{\sim}58,400/\allowbreak \ensuremath{\sim}1,200, 2,674/102 \\
\bottomrule
\end{tabular}
\normalsize
\end{table}

\subsection{AI workflow}\label{ai-workflow}

To try to better understand the AI test writing workflow used in Quay and AWX we (i) analysed the pattern of commits, (ii) contacted the authors.

\subsubsection{Commit analysis}\label{commit-analysis}

\begin{figure}
\centering
\includegraphics[width=\linewidth,keepaspectratio]{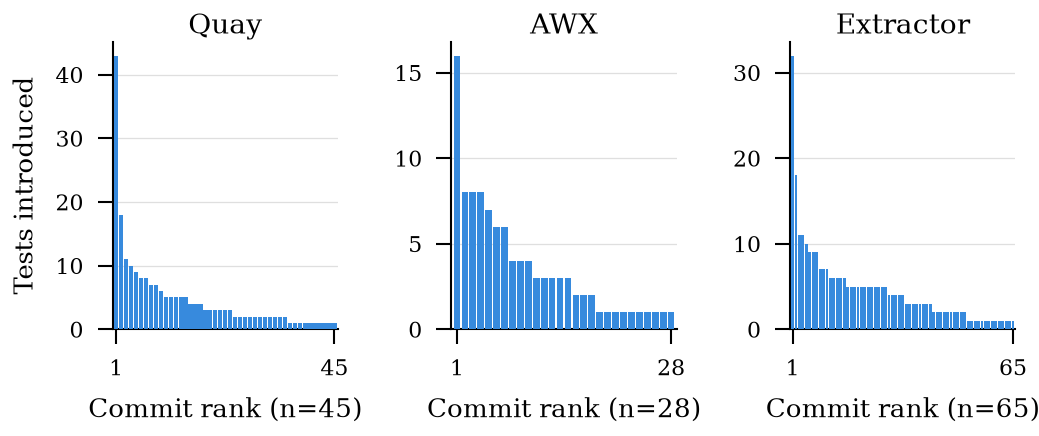}
\caption{Tests introduced per individual commit, sorted in descending order, for Quay, AWX, and Extractor.}\label{fig:tests-per-commit-distributions}
\end{figure}

In Quay and AWX PRs multiple commits are merged. However, the individual pre-merge commits can be recovered from the GitHub API (via \texttt{refs/pull/\textless{}N\textgreater{}/head} and \texttt{\seqsplit{git\ blame}}) and so we are able to analyse these. Both the Quay and AWX commits containing AI written tests almost always bundle a feature's implementation and unit tests together. Figure \ref{fig:tests-per-commit-distributions} shows that each commit typically contains multiple tests, similarly to Extractor.

In summary, the commit analysis indicates that the AI tests were generally developed incrementally alongside new production code.

\subsubsection{Communication with code authors}\label{communication-with-code-authors}

We contacted the code authors asking for information about the workflow they used for AI-authored tests. A Quay maintainer confirmed Claude Code produced these PRs largely autonomously: Claude was instructed to take a JIRA ticket number as input and to carry out the entire implementation, with explicit instructions to write both unit and integration tests along the way\footnote{Quay's /create-plan-from-issue command, e.g.~as of commit 92eddf2: \url{https://github.com/quay/quay/blob/92eddf2dbc266eeed2d138c6dcaa7fe38c5f7479/.claude/commands/create-plan-from-issue.md}}.

We did not receive a response from the AWX developers that we contacted, so the human input to the AI test writing is unclear. In particular, the level of detail given in the instructions to Claude and the amount of human review / re-writing of the tests are both unknown.

\subsection{Coverage-guided mutation results}\label{coverage-guided-mutation-results}

\begin{table}[tb]
\footnotesize
\centering
\caption{AST mutation testing results by corpus, coverage-based protocol.}\label{tab:extras-killrates}
\begin{tabular}{@{}
  >{\raggedright\arraybackslash}p{(\linewidth - 12\tabcolsep) * \real{0.1109}}
  >{\raggedright\arraybackslash}p{(\linewidth - 12\tabcolsep) * \real{0.1694}}
  >{\raggedright\arraybackslash}p{(\linewidth - 12\tabcolsep) * \real{0.1694}}
  >{\raggedright\arraybackslash}p{(\linewidth - 12\tabcolsep) * \real{0.1109}}
  >{\raggedright\arraybackslash}p{(\linewidth - 12\tabcolsep) * \real{0.1281}}
  >{\raggedright\arraybackslash}p{(\linewidth - 12\tabcolsep) * \real{0.1358}}
  >{\raggedright\arraybackslash}p{(\linewidth - 12\tabcolsep) * \real{0.1754}}
@{}}
\toprule
\begin{minipage}[b]{\linewidth}\raggedright
Corpus
\end{minipage} & \begin{minipage}[b]{\linewidth}\raggedright
Tests (SCORED)
\end{minipage} & \begin{minipage}[b]{\linewidth}\raggedright
Scored mutants
\end{minipage} & \begin{minipage}[b]{\linewidth}\raggedright
Killed
\end{minipage} & \begin{minipage}[b]{\linewidth}\raggedright
Survived
\end{minipage} & \begin{minipage}[b]{\linewidth}\raggedright
Kill rate
\end{minipage} & \begin{minipage}[b]{\linewidth}\raggedright
95\% CI (Wilson)
\end{minipage} \\
\midrule
Quay & 128 & 1,825 & 817 & 1,008 & 44.8\% & {[}42.5\%, 47.1\%{]} \\
AWX & 50 & 665 & 391 & 274 & 58.8\% & {[}55.0\%, 62.5\%{]} \\
\bottomrule
\end{tabular}
\normalsize
\end{table}

\begin{table}[tb]
\footnotesize
\centering
\caption{One-sided 95\% non-inferiority bounds on the coverage-guided kill-rate, by corpus.}\label{tab:extras-killrates-bounds}
\begin{tabular}{@{}
  >{\raggedright\arraybackslash}p{(\linewidth - 4\tabcolsep) * \real{0.3217}}
  >{\raggedleft\arraybackslash}p{(\linewidth - 4\tabcolsep) * \real{0.3391}}
  >{\raggedleft\arraybackslash}p{(\linewidth - 4\tabcolsep) * \real{0.3391}}
@{}}
\toprule
Corpus & vs.~Django & vs.~Pandas \\
\midrule
Quay & 7.8\% & 3.8\% \\
AWX & 0\% & 0\% \\
Extractor & 5.9\% & 2.0\% \\
\bottomrule
\end{tabular}
\normalsize
\end{table}

Table \ref{tab:extras-killrates} shows the measured kill-rates for Quay and AWX using coverage-guided mutation, and Table \ref{tab:extras-killrates-bounds} shows the corresponding non-inferiority bounds relative to Django and Pandas. The latter rules out Quay's kill rate being worse than Django/Pandas by more than 7.8\%/3.8\%, at 95\% confidence, and rules out AWX's kill rate being worse than Django or Pandas at all. The Quay bounds are larger than for Extractor but this is at least partly because of the smaller sample size, which increases the uncertainty in the measured critical rate and so increases the bound. Bearing in mind the discussion in \S\ref{sec:interpret-kill-rate}, the data suggests that the test suites of all three AI-corpora provide similar levels of fault protection to Django and Pandas.

\subsection{Qualitative test-quality results}\label{qualitative-test-quality-results}

\begin{table}[tb]
\footnotesize
\centering
\caption{Per-axis severity breakdown, by corpus.}\label{tab:extras-severity}
\begin{tabular}{@{}
  >{\raggedright\arraybackslash}p{(\linewidth - 14\tabcolsep) * \real{0.1481}}
  >{\raggedleft\arraybackslash}p{(\linewidth - 14\tabcolsep) * \real{0.1391}}
  >{\raggedleft\arraybackslash}p{(\linewidth - 14\tabcolsep) * \real{0.1524}}
  >{\raggedleft\arraybackslash}p{(\linewidth - 14\tabcolsep) * \real{0.1016}}
  >{\raggedleft\arraybackslash}p{(\linewidth - 14\tabcolsep) * \real{0.1245}}
  >{\raggedleft\arraybackslash}p{(\linewidth - 14\tabcolsep) * \real{0.0982}}
  >{\raggedleft\arraybackslash}p{(\linewidth - 14\tabcolsep) * \real{0.1344}}
  >{\raggedleft\arraybackslash}p{(\linewidth - 14\tabcolsep) * \real{0.1016}}
@{}}
\toprule
\begin{minipage}[b]{\linewidth}\raggedright
\end{minipage} & \begin{minipage}[b]{\linewidth}\raggedleft
Intent-fidelity
\end{minipage} & \begin{minipage}[b]{\linewidth}\raggedleft
Assertion strength
\end{minipage} & \begin{minipage}[b]{\linewidth}\raggedleft
Coupling
\end{minipage} & \begin{minipage}[b]{\linewidth}\raggedleft
Anti-pattern
\end{minipage} & \begin{minipage}[b]{\linewidth}\raggedleft
Setup
\end{minipage} & \begin{minipage}[b]{\linewidth}\raggedleft
Purpose-stated
\end{minipage} & \begin{minipage}[b]{\linewidth}\raggedleft
Cohesion
\end{minipage} \\
\midrule
\multicolumn{8}{>{\raggedright\arraybackslash}p{\linewidth-2\tabcolsep}}{\textbf{Quay (n=214, critical=8 (3.7\%), moderate=39 (18.2\%))}}\\
Critical & 7 & 4 & 0 & 2 & 5 & 0 & 0 \\
Moderate & 17 & 27 & 2 & 5 & 4 & 4 & 1 \\
Minor & 0 & 0 & 0 & 3 & 1 & 3 & 0 \\
Info & 0 & 0 & 13 & 0 & 0 & 45 & 0 \\
Critical+\allowbreak Moderate & 24 & 31 & 2 & 7 & 9 & 4 & 1 \\
\midrule
\multicolumn{8}{>{\raggedright\arraybackslash}p{\linewidth-2\tabcolsep}}{\textbf{AWX (n=102, critical=3 (2.9\%), moderate=15 (14.7\%))}}\\
Critical & 3 & 1 & 0 & 2 & 3 & 2 & 0 \\
Moderate & 2 & 13 & 2 & 0 & 1 & 5 & 0 \\
Minor & 0 & 0 & 0 & 0 & 0 & 1 & 0 \\
Info & 0 & 0 & 16 & 0 & 0 & 31 & 0 \\
Critical+\allowbreak Moderate & 5 & 14 & 2 & 2 & 4 & 7 & 0 \\
\bottomrule
\end{tabular}
\normalsize
\end{table}

\begin{table}[tb]
\footnotesize
\centering
\caption{One-sided 95\% non-inferiority bounds on the qualitative-rubric critical-defect rate, by corpus. Tests are attributable to Opus 4.6 and later.}\label{tab:extras-severity-bounds}
\begin{tabular}{@{}
  >{\raggedright\arraybackslash}p{(\linewidth - 4\tabcolsep) * \real{0.3217}}
  >{\raggedleft\arraybackslash}p{(\linewidth - 4\tabcolsep) * \real{0.3391}}
  >{\raggedleft\arraybackslash}p{(\linewidth - 4\tabcolsep) * \real{0.3391}}
@{}}
\toprule
Corpus & vs.~Django & vs.~Pandas \\
\midrule
Quay & 4.8\% & 5.6\% \\
AWX & 4.6\% & 5.6\% \\
Extractor & 2.3\% & 3.2\% \\
\bottomrule
\end{tabular}
\normalsize
\end{table}

Table \ref{tab:extras-severity} summarises the results of the qualitative evaluations of Quay and AWX, while Table \ref{tab:extras-severity-bounds} shows bootstrapped one-sided non-inferiority bounds for their critical rates relative to Django and Pandas. These bounds rule out Quay's critical rate exceeding Django/Pandas by more than 4.8\%/5.6\% and AWX's by more than 4.6\%/5.6\%.

Also included in the table are the Extractor values from \S\ref{sec:qualitative}. It can be seen that the Quay and AWX bounds are somewhat larger than the bounds for Extractor. This is partly because of the smaller sample sizes for Quay, and especially AWX, which increases the uncertainty in the measured critical rate and so increases the bound\footnote{Keeping the percentage of critical samples fixed then for a population of 298 tests (matching Extractor) the AWX bounds decrease to 3.5\%/4.4\% from 4.6\%/5.6\%, and the Quay bounds decrease to 4.3\%/5.4\%}. Nevertheless, the relative number of critical tests for Quay and AWX does seem slightly higher than for Extractor: 2.9\% (3/102) critical AWX tests and 3.7\% (8/214) Quay tests vs 2.0\% (6/298) for Extractor. However the absolute number of tests flagged as critical is small (3/8/6) and so this comparison should be read cautiously since changes in one or two tests would be enough to make the numbers for all three comparable.

\subsection{Opus 4.5 and earlier}\label{sec:opus45-earlier}

\begin{table}[tb]
\scriptsize
\centering
\caption{Per-axis severity breakdown for Quay tests attributable to Opus 4.5 and earlier.}\label{tab:extras-severity-opus45}
\begin{tabular}{@{}
  >{\raggedright\arraybackslash}p{(\linewidth - 14\tabcolsep) * \real{0.1481}}
  >{\raggedleft\arraybackslash}p{(\linewidth - 14\tabcolsep) * \real{0.1391}}
  >{\raggedleft\arraybackslash}p{(\linewidth - 14\tabcolsep) * \real{0.1524}}
  >{\raggedleft\arraybackslash}p{(\linewidth - 14\tabcolsep) * \real{0.1016}}
  >{\raggedleft\arraybackslash}p{(\linewidth - 14\tabcolsep) * \real{0.1245}}
  >{\raggedleft\arraybackslash}p{(\linewidth - 14\tabcolsep) * \real{0.0982}}
  >{\raggedleft\arraybackslash}p{(\linewidth - 14\tabcolsep) * \real{0.1344}}
  >{\raggedleft\arraybackslash}p{(\linewidth - 14\tabcolsep) * \real{0.1016}}
@{}}
\toprule
\begin{minipage}[b]{\linewidth}\raggedright
\end{minipage} & \begin{minipage}[b]{\linewidth}\raggedleft
Intent-fidelity
\end{minipage} & \begin{minipage}[b]{\linewidth}\raggedleft
Assertion strength
\end{minipage} & \begin{minipage}[b]{\linewidth}\raggedleft
Coupling
\end{minipage} & \begin{minipage}[b]{\linewidth}\raggedleft
Anti-pattern
\end{minipage} & \begin{minipage}[b]{\linewidth}\raggedleft
Setup
\end{minipage} & \begin{minipage}[b]{\linewidth}\raggedleft
Purpose-stated
\end{minipage} & \begin{minipage}[b]{\linewidth}\raggedleft
Cohesion
\end{minipage} \\
\midrule
\multicolumn{8}{>{\raggedright\arraybackslash}p{\linewidth-2\tabcolsep}}{\textbf{Quay (n=233, critical=24 (10.3\%), moderate=67 (28.8\%))}}\\
Critical & 17 & 14 & 0 & 11 & 13 & 3 & 0 \\
Moderate & 22 & 52 & 5 & 8 & 7 & 7 & 3 \\
Minor & 0 & 2 & 0 & 1 & 4 & 4 & 0 \\
Info & 0 & 0 & 9 & 0 & 0 & 18 & 0 \\
Critical+\allowbreak Moderate & 39 & 66 & 5 & 19 & 20 & 10 & 3 \\
\bottomrule
\end{tabular}
\normalsize
\end{table}

We excluded tests written by Claude Opus 4.5 and earlier because they seem to be of significantly lower quality than the tests written by later models. Table \ref{tab:extras-severity-opus45} shows the qualitative evaluation of a sample of Quay tests written by Opus 4.5 and earlier. Observe that 24 out of 233 tests are flagged as critical i.e.~10.3\%. For comparison, Table \ref{tab:extras-severity} shows that for Quay tests written by Opus 4.6 and later only 8 out of 214 tests are flagged critical i.e.~3.7\%. The available AWX tests are all written by Opus 4.6 and later models, similarly for Extractor.

\section{Discussion}\label{discussion}

\subsection{Are Django/Pandas tainted by AI-authored tests?}\label{are-djangopandas-tainted-by-ai-authored-tests}

We use Django/Pandas as human-authored baselines against which to compare the AI-authored Extractor tests. Django/Pandas are both actively maintained and still accepting commits today, a period in which
AI-assisted coding has become more common. Both projects have AI-contribution policies, adopted in January 2026 \cite{django2026aipolicy, pandas2026aipolicy} which require disclosure but neither specifies a machine-checkable
format or a commit-message location for it. Searching every pull request's body directly for
AI-disclosure reveals 20 pandas PRs (26 tests) and 12 django PRs (12 tests) . These PRs all date from January 2026 or later. Cross-referencing these against the test sample used in our evaluation shows that the samples draw overwhelmingly from pre-2026 commits (only 2 of 351 Django rows post-date the AI-contribution policy). We find that recomputing each corpus' historical-revert rate with the disclosed AI tests excluded changes the reported scores by less than 0.1\%. As well as explicit disclosures, we also searched the commit text of all the tests used in our sample for keywords and phrases that might suggest AI-authorship and found none.

\subsection{Are the AI-authored tests tainted by the AI's training data?}\label{are-the-ai-authored-tests-tainted-by-the-ais-training-data}

In this study we seek to evaluate the quality of tests which are AI-authored during code development. One evaluation approach would be to take a public code corpus, strip of its (likely human-authored) tests, and then ask an AI to create tests for the resulting corpus. However, there are two main problems with this approach. Firstly, the AI tests are written post code development, not during it. Secondly, public code, including its existing tests, is almost certainly incorporated in the training data of any frontier code-writing AI. That is, the AI has already had sight of the existing tests before they are stripped and this runs the risk of confounding any evaluation of AI test authorship\footnote{\cite{evolution2026residual}'s study of LLM test generation under code evolution likely gives a concrete illustration of this. 99\% of AI tests that fail once the code under test has genuinely changed still pass against the original, pre-change version, while exercising the same, now-changed region. The AI tests are observed track what the code used to do rather than what it currently does, which is what memorisation of that code, drawn from the public Project CodeNet corpus, might well look like.}.

In this study we avoid both of these concerns. The Extractor code was developed from scratch in a private environment, and has never been published. The tests are therefore authored during code development, rather than after, and it is impossible for the AI model to have been trained on this code or otherwise internalised it into its weights. Every Extractor test scored in this report reflects the model's own test-writing behaviour, not recall of prior exposure.

\subsubsection{What does ``human-authored'' mean here?}\label{what-does-human-authored-mean-here}

For Django and Pandas, ``human-authored'' means authored by a group of humans with the processes a mature open-source project runs. Namely, a contributor writes the test, a maintainer distinct from that contributor reviews the pull request before merge, and, for many of the bugs sampled here, an external user found and reported the defect the test exists to catch. The AI-authored tests in Extractor have no analogue to any of this. They were written by one AI agent with one human directing it: no independent reviewer, no external user base, and no detailed test-design or review guidance from that human either (see Additional Material \S \ref{sec:prompting}). It is notable that the AI-authored tests in Extractor match the human-authored Django and Pandas baselines despite lacking all of that scaffolding.

\subsection{Limitations}\label{limitations}

\subsubsection{Scope: one language, one model family}\label{scope-one-language-one-model-family}

The findings in this report analyse Python code authored by one AI model family (Claude Sonnet/Opus). We make no claim that these findings generalise to other coding languages or AI models (e.g.~GPT Codex). In particular, we expect the results to be sensitive to the language used. Python appears to be Claude's preferred language in the sense that when Claude organically writes tools for its own use they are written in Python. Claude is therefore probably especially well trained in Python, and its Python code-writing performance may be better than for other languages.

\subsubsection{Low base rates limit statistical power}\label{low-base-rates-limit-statistical-power}

As already noted, because the measured test quality is high even evaluating hundreds of tests per corpus still only turns up a small absolute number of tests flagged as having potential flaws. These small counts mean that the reported rates are sensitive to small changes in the number of flagged tests. That sensitivity matters less, however, for our main conclusion that all three corpora have low overall defect rates (a few per cent) and that AI-written tests are not weaker than human-written ones on this corpus. This conclusion only needs that the failure rate be reliably small, not pinned down to a precise value. A few hundred tests per corpus is enough to establish that with reasonable confidence, even if it is not enough to adjudicate a one- or two-point gap between corpora.

\section{Conclusion}\label{conclusion}

We score Claude-written tests against two human-written corpora under four independent methods:
historical-revert testing, per-commit AST mutation testing, coverage-guided mutation testing, and a
seven-axis qualitative rubric. This is at a larger combined scale than any prior individual-test
evaluation we are aware of. On every method, the AI-written tests are broadly comparable to the two
human-written corpora, a result that runs against the more pessimistic picture from prior
AI-test-generation studies based on compile/pass rates.

\bibliography{references}

\clearpage

\section{Additional Material}\label{additional-material}

\subsection{Historical-revert evaluation procedure}\label{historical-revert-evaluation-procedure}

See Figure \ref{fig:historical-revert-pseudocode} for pseudo-code of the historical-revert evaluation protocol used. This is executed once per candidate \texttt{\seqsplit{(c,\ t,\ F)}}, where \texttt{\seqsplit{c}} is
the test's introducing (post-fix) commit, \texttt{\seqsplit{t}} is the test, and \texttt{\seqsplit{F}} is the set
of production files touched by \texttt{\seqsplit{c}}'s diff. Build-environment provisioning (era-appropriate compiler/interpreter/dependency
pins) is treated as a precondition of \texttt{\seqsplit{RUN}} and omitted from the figure for brevity. The coverage.py tool used to evaluate code coverage is described below

\FloatBarrier

\begin{figure*}[tb]
\footnotesize
\begin{verbatim}
Algorithm GROUP-B-EVAL(commit c, test t, production files F)
    checkout(worktree, c)                        // post-fix state
    if RUN(t) fails:
        return BASELINE_FAILS                     // environment/build issue, excluded

    for f in F:
        WRITE(f, content_of(f, parent(c)))         // revert to pre-fix content

    if RUN(t) fails:
        verdict <- EFFECTIVE                        // test catches the historical bug
    else:
        verdict <- REACHABILITY-CHECK(t, F)          // passes even without the fix — see below

    for f in F:
        WRITE(f, content_of(f, c))                 // restore post-fix state
    return verdict

Sub-procedure REACHABILITY-CHECK(test t, production files F)
    trace <- RUN(t, instrumented with coverage.py --branch, scoped to F)
    changed <- the revert diff's '+' lines (blank lines excluded — see below)
    if any line/branch in `changed` shows as executed in `trace`:
        return SURVIVED           // this run genuinely reached the changed code, just didn't discriminate
    else:
        return UNREACHED          // this run never executed the changed code at all
\end{verbatim}
\caption{Historical-revert evaluation procedure (pseudo-code).}\label{fig:historical-revert-pseudocode}
\end{figure*}

\subsection{Qualitative rubric agent instructions}\label{qualitative-rubric-agent-instructions}

Figure \ref{fig:qualitative-rubric-instructions} shows the operative instruction text given to the rubric scoring agents, stripped of per-dataset specifics (sample file paths, repo locations, batch sizes) i.e.~this covers the eval process and scoring rules themselves, which do not change between datasets. Each test is scored by its own agent (there is no batching). Scoring is a two-stage process: a blind Stage A pass scores all seven axes from the test's source alone, and a Stage B pass then re-examines the two axes mutation evidence bears on most directly (assertion strength, setup/fixture proportionality) in light of that test's own mutants.

After scoring, the defects in each test are evaluated for severity. The four severity tiers used are given in Table \ref{tab:severity-definitions} and the instructions given to the first pass agent are shown in Figure \ref{fig:qualitative-severity1-instructions}. Tests categorised as Critical severity in the first pass are double checked by a panel of three independent agents, each of which re-derives whether the test deserves a ``critical'' severity tier for its test-quality defect. The instructions given to the panelists are shown in Figure \ref{fig:qualitative-severity2-instructions}. If a majority agree on Critical then that category is kept, otherwise the severity is downgraded to Moderate.

Table \ref{tab:stage-ab-changes} shows the changes in scoring between Stage A and Stage B for tests categorised with Critical or Moderate severity. It can be seen that changes have relatively more impact on the Critical than on the Moderate tests. Table \ref{tab:stage-ab-changes2} shows the corresponding breakdown by rubric scoring axis.

\begin{figure*}[bh]
\fontsize{6}{7.2}\selectfont
\noindent\fbox{\begin{minipage}{\dimexpr\linewidth-2\fboxsep-2\fboxrule\relax}
\emph{1) Reading order (given before any scoring):}

Read every candidate test's ENTIRE
file from line 1 -- module docstring, class docstring, method docstring/comment -- before
scoring axis 6. Never rely on just the method body, or on an excerpt from another document:
this is the single most common way a prior pass got a purpose-clarity score wrong.

\emph{2) The seven axes and how to score each:}

\textbf{Axis 1 -- intent-fidelity.} Does the assertion actually check what the docstring/name/
comment claims? \texttt{\seqsplit{MATCH}} / \texttt{\seqsplit{PARTIAL}} (checks something related but narrower or looser than
stated) / \texttt{\seqsplit{MISMATCH}} (passes or fails for a reason unrelated to the stated claim). Ground
this by tracing what the assertion can and cannot distinguish, not by reading the name and
stopping there. Write your reasoning to a notes field -- a bare enum value with no notes
leaves no way to check \emph{why} a verdict was reached, which blocks auditing disagreements
between passes.

\textbf{Axis 2 -- assertion strength.} \texttt{\seqsplit{STRICT}} / \texttt{\seqsplit{MODERATE}} / \texttt{\seqsplit{WEAK}} / \texttt{\seqsplit{n/a}}. This is a precision
judgment, not a syntactic one: STRICT means the assertion is at the ceiling of precision
available for what's actually being claimed, without introducing fragility the claim doesn't
call for -- literal equality when the true value is a fixed fact, a threshold when the true
value is inherently a range that legitimately varies, existence when the claim is inherently
about existence. \texttt{\seqsplit{MODERATE}}/\texttt{\seqsplit{WEAK}} is only a valid score if you can present a concrete,
robustness-preserving change that reaches STRICT. If no such change exists -- because
tightening it would introduce fragility, or duplicate a sibling test's job -- promote to
STRICT instead of leaving it as an unfalsifiable ``could be better in theory.'' Do not count
the number of assertions as a proxy for strength.

\textbf{Axis 3 -- behavior/implementation coupling.} \texttt{\seqsplit{BEHAVIOR}} (asserts an externally observable
outcome; survives a legitimate refactor) / \texttt{\seqsplit{JUSTIFIED-MECHANISM}} (couples to an
implementation detail, but that detail \emph{is} the fix under test, so the coupling is
appropriate) / \texttt{\seqsplit{UNJUSTIFIED-COUPLING}} (asserts an incidental implementation detail that would
break under a behavior-preserving refactor and would teach a future maintainer nothing). Write
your reasoning to a notes field, same requirement and rationale as axis 1 above; a bare enum value with no notes gives a later rescore nothing to check disagreements against.

\textbf{Axis 4 -- anti-pattern scan.} A checklist, not a scale. A single-select field -- pick the
one category that most specifically owns the finding, never a comma-joined list of several.
Only use these exact strings, nothing else: \texttt{\seqsplit{none}} / \texttt{\seqsplit{dead\ lever}} / \texttt{\seqsplit{assertion\ masking\ (coverage-real)}} (verify by tracing the actual implementation, don't guess from syntax) /
\texttt{\seqsplit{assertion\ masking\ (diagnostic-only)}} (the sequential assertions provably share one
underlying line/call) / \texttt{\seqsplit{over-broad\ mock}} / \texttt{\seqsplit{swallowed\ exception}} / \texttt{\seqsplit{log-instead-of-assert}}
/ \texttt{\seqsplit{tautology}} (cannot fail under any input) / \texttt{\seqsplit{skip/xfail\ masking}}. If a test genuinely
exhibits more than one, name the single most severe/specific one here and mention the others
in the notes field -- the column itself must stay one of the exact strings above.

\textbf{Axis 5 -- setup/fixture proportionality.} Output is \texttt{\seqsplit{OK}} / \texttt{\seqsplit{DEFECTIVE}} / \texttt{\seqsplit{n/a}} -- exactly
one of these three strings. Two sub-questions inform the verdict: does Arrange build only
what's needed (excess/bloat), and does the fixture actually construct the \emph{scenario} the test
claims to exercise, or does it accidentally construct a different one that happens to produce
the same superficial outcome for an unrelated reason? Either failure mode scores \texttt{\seqsplit{DEFECTIVE}}
-- say which one, and why, in the notes field. The second is the more severe failure mode: no
mutation of the actual logic under test can ever be caught if the defect is upstream in
Arrange.

\textbf{Axis 6 -- purpose stated.} Does \emph{some} source state what the test verifies and why? Two
independent things determine the verdict: whether a qualifying source exists at all, and --
if so -- whether a reader encounters it \textbf{in the file itself} (docstring/comment at any
granularity) or \textbf{only in VCS history} (the introducing commit message, with nothing
recovered in-file). These are not interchangeable -- a reader who never runs \texttt{\seqsplit{git\ log}} on this
file never sees the second kind -- and collapsing them into one CLEAR value is a confirmed source of disagreement between scoring passes. Record which family explicitly::

\begin{itemize}
\tightlist
\item
  \texttt{\seqsplit{CLEAR\_INFILE\_SOURCE\ (per-test)}} / \texttt{\seqsplit{CLEAR\_INFILE\_SOURCE\ (class-level)}} / \texttt{\seqsplit{CLEAR\_INFILE\_SOURCE\ (module-level)}} / \texttt{\seqsplit{CLEAR\_INFILE\_SOURCE\ (external)}} (via
  an in-file reference to an issue tracker) -- any of these require an actual
  docstring/comment/in-file trace, not a method name that ``sounds right'' (a name is not a
  stated source at any tier).
\item
  \texttt{\seqsplit{CLEAR\_COMMIT\_TEXT\ (per-test)}} (the introducing commit is
  specifically about this one test) / \texttt{\seqsplit{CLEAR\_COMMIT\_TEXT\ (shared)}} (the commit explains the
  purpose but is a broader/batch commit not calling this test out individually) -- use this pair
  only when nothing in-file states the purpose at any granularity. A commit message is a real,
  legitimate source -- don't invent a stricter bar than the rubric already has -- but it is measurably weaker than an in-file one and must be recorded as such, never silently folded into \texttt{\seqsplit{CLEAR\_INFILE\_SOURCE}}.
\item
  \texttt{\seqsplit{PARTIAL}} (something exists -- in-file or commit -- but under-explains) / \texttt{\seqsplit{WRONG}} (a source exists, is
  specific, and is factually incorrect) / \texttt{\seqsplit{NONE}}.
\end{itemize}

Check \textbf{all} of: in-file docstring/comment at
every granularity, the introducing commit message, and whether an external tracker is
reachable via an in-file reference -- before concluding anything, and before picking which
family applies.

\textbf{Axis 7 -- cohesion (one thing per test).} Does the test exercise a single cohesive
behavioral claim (\texttt{\seqsplit{SINGLE}}), even via multiple assertions that are all facets of that one
claim, or does it bundle two or more genuinely independent claims into one method
(\texttt{\seqsplit{BUNDLED}})? For parametrized tests, each case checking the same claim under different inputs
is \texttt{\seqsplit{SINGLE}}, not \texttt{\seqsplit{BUNDLED}}.

\emph{3) Scoring:}

Score all seven axes above from the test's source (and, for axis 6, its commit) alone.

\emph{3) Stage A -- blind scoring}
Score all seven axes above from the test's source (and, for axis 6, its commit) alone.

Do not open, list, or otherwise consult any mutation-testing output for this test -- no coverage\_guided\_mutation\_study results, no ast\_mutation\_study results, no kill rate, no mutant list. You may still perform source-grounded live verification (run the test as it stands, or hand-construct your own corrupted/mutated state and run the real assertion against it) -- that is your own reasoning about the test, not a lookup into either mutation study's corpus, and does not leak Stage-B information.

Write Stage A's verdicts to axis1\_blind \ldots{} axis7\_blind, with per-axis notes. These columns are frozen once written -- Stage B never edits them.

\emph{4) Stage B -- mutation-informed refinement}
Given: Stage A's own verdicts/notes, plus this test's candidate mutants. Re-examine axis 2 and axis 5 in light of them -- these are the two axes a concrete survived/killed mutant most directly bears on -- but any axis may be revised if the evidence genuinely changes the picture (e.g.~a survived mutant in dead code bears on axis 4's dead lever, not axis 2).

Candidate-mutant sourcing, in priority order:

\begin{enumerate}
\def\labelenumi{(\roman{enumi})}
\tightlist
\item
  Primary: this corpus's ast\_mutation\_study results, keyed by commit hash, already pinned to the test's own introducing commit.
\item
  Supplementary only: this corpus's coverage\_guided\_mutation\_study results -- generated against current HEAD, not necessarily this test's own commit, so its killed/survived label is never trusted directly.
\item
  If neither source has any mutants for this test, copy Stage A's verdicts forward unchanged and stop -- don't fabricate mutation evidence for a test that has none.
\end{enumerate}

Verification is mandatory and must happen at the test's own pinned commit -- never against the live working tree. For each candidate mutant relied on: create a dedicated, throwaway git worktree at the test's own commit (never a shared worktree another agent might be using concurrently), hand-patch the mutation into the pinned source, rerun the real test there, and use that pass/fail result -- not any study's recorded label -- as the evidence. Before treating a patched-in survivor as a real weakness, confirm the mutated line is actually reachable by this specific test -- most survivors are unreachable-by-this-test noise (dead code, another test's path, a downstream guard already masking the outcome), not genuine gaps. Revert the hand-patch (or discard the worktree) once done; never leave a mutated file checked in.

Write to axis1\_refined \ldots{} axis7\_refined, plus per-axis refine notes for anything that changed, and a single refine\_delta = yes/no. Also record mutant\_source.

Sanity expectation, not a target to game: mutation evidence should essentially never move a verdict stricter than blind Stage A -- only weaker or unchanged. A run showing a meaningful rate of stricter changes should be treated as a likely leakage signal (e.g.~accidentally reading the live working tree instead of the pinned worktree), not a new finding to report at face value.

\emph{4) Output columns:}
corpus, test\_id, file, commit\_hash, axis1\_blind, axis1\_blind\_notes, axis2\_blind, axis2\_blind\_notes, axis3\_blind, axis3\_blind\_notes, axis4\_blind, axis4\_blind\_notes, axis5\_blind, axis5\_blind\_notes, axis6\_blind, axis6\_blind\_notes, axis7\_blind, axis7\_blind\_notes, axis1\_refined, axis1\_refined\_notes, axis2\_refined, axis2\_refined\_notes, axis3\_refined, axis3\_refined\_notes, axis4\_refined, axis4\_refined\_notes, axis5\_refined, axis5\_refined\_notes, axis6\_refined, axis6\_refined\_notes, axis7\_refined, axis7\_refined\_notes, refine\_delta, mutant\_source, flag

flag = yes if any axisN value is non-ideal, else no. Ideal values: axis1=MATCH, axis2=STRICT, axis3=BEHAVIOR, axis4=none, axis5=OK, axis6=CLEAR\_INFILE\_SOURCE (\emph{) at any granularity, axis7=SINGLE. CLEAR\_COMMIT\_TEXT (}) is not ideal and must flag -- same tier as PARTIAL.
\end{minipage}}
\caption{Qualitative rubric agent instructions.}\label{fig:qualitative-rubric-instructions}
\end{figure*}

\footnotesize

\begin{table*}[tb]
\footnotesize
\centering
\caption{Severity tier definitions for non-ideal rubric verdicts.}\label{tab:severity-definitions}
\begin{tabular}{@{}
  >{\raggedright\arraybackslash}p{(\linewidth - 2\tabcolsep) * \real{0.4149}}
  >{\raggedright\arraybackslash}p{(\linewidth - 2\tabcolsep) * \real{0.5851}}
@{}}
\toprule
\begin{minipage}[b]{\linewidth}\raggedright
Severity
\end{minipage} & \begin{minipage}[b]{\linewidth}\raggedright
Definition
\end{minipage} \\
\midrule
Critical & A genuine behavioural defect: axis-1 \texttt{\seqsplit{MISMATCH}}, axis-5 \texttt{\seqsplit{DEFECTIVE}} with no mitigation, or an axis-2 strength \texttt{\seqsplit{WEAK}} or \texttt{\seqsplit{MODERATE}} proven to have missed a real mutation. \\
Moderate & A real, presentable gap, with no proof yet of an actual missed defect. \\
Minor & Diagnostic-only assertion masking, a tautology, or an axis-6 \texttt{\seqsplit{WRONG}}/\texttt{\seqsplit{PARTIAL}} docstring that does not affect the test's own correctness. \\
Informational & axis-3 \texttt{\seqsplit{JUSTIFIED-MECHANISM}}, axis-6 \texttt{\seqsplit{CLEAR\_COMMIT\_TEXT}} when it's the \\
row's only flagged reason; any finding whose own note calls the deviation borderline/trivial. & \\
\bottomrule
\end{tabular}
\normalsize
\end{table*}

\normalsize

\begin{figure*}[tb]
\footnotesize
\noindent\fbox{\begin{minipage}{\dimexpr\linewidth-2\fboxsep-2\fboxrule\relax}
You are proposing a severity tier for already-scored, already-flagged qualitative test-quality rubric rows from the {[}corpus name{]} corpus. Each file you've been given is a standalone CSV -- one header row plus one data row -- already fully scored across all seven axes (schema v2), with flag = yes.

Your job: read each row's own already-recorded notes and propose a severity tier for the row as a whole. Do not re-run anything, do not open source files, and do not invent new evidence -- work entirely from what is already written in the row's own notes, paying particular attention to whether a finding's verification was a genuine live hand-patch-and-rerun (look for phrases like ``hand-verified,'' ``reran pytest,'' ``patched \ldots{} rerun \ldots{} FAILED/PASSED'') versus source-only reasoning (no mutation evidence available, verdict copied forward, ``NO\_CANDIDATES'').

Important: when you report the test's identity, copy the value verbatim from the row's own test\_id column -- never derive it from the filename or shorten it.

\emph{Severity tiers}

Severity tiers are: critical, moderate, minor, informational.

Critical is narrow -- reachable from only three findings, and nothing else qualifies no matter how severe it reads:

axis1\_refined = MISMATCH, and the row's own notes record a live check (a test run, or a hand-patch followed by a rerun) that concretely demonstrates the mismatch.
axis5\_refined = DEFECTIVE for the ``wrong scenario'' reason (not just excess/bloat setup), and the row's notes record a live check proving it.
axis2\_refined = WEAK or MODERATE, and the row's Stage B notes show that the live check itself constructed the corrupted or mutated state, reran the real test, and the existing assertion still passed.
If the verification behind an axis-1/2/5 finding was source-only, based on NO\_CANDIDATES, or simply copied forward from Stage A without live evidence, it cannot be critical -- cap it at moderate. Findings on axis 3, axis 4, axis 6, or axis 7 can never reach critical on their own.

Moderate: a real, well-reasoned defect that doesn't meet the live-verified bar above -- axis-2 WEAK/MODERATE backed only by source-level reasoning; axis-1 PARTIAL; axis-5 DEFECTIVE without live proof; axis-4 anti-patterns that materially reduce the test's value (``assertion masking (coverage-real),'' ``over-broad mock,'' ``swallowed exception,'' ``tautology,'' ``skip/xfail masking''); axis-7 BUNDLED where the bundling genuinely obscures which claim failed.

Minor: axis-4 findings where a real assertion still exists elsewhere and the practical cost is low (``dead lever,'' ``assertion masking (diagnostic-only),'' ``log-instead-of-assert''); axis-5 DEFECTIVE for excess/bloat only, not wrong-scenario; axis-6 PARTIAL or WRONG.

Informational: axis-3 JUSTIFIED-MECHANISM (by the axis's own definition, this coupling is appropriate -- it flags mechanically but isn't a real defect); axis-6 CLEAR\_COMMIT\_TEXT when it is the row's only flagged reason; any finding whose own note describes the deviation as borderline or trivial.

The CLEAR\_COMMIT\_TEXT rule (this overrides anything above that would imply otherwise): axis6\_refined = CLEAR\_COMMIT\_TEXT never elevates severity -- a commit message is a real, legitimate purpose source, just weaker than an in-file one, and is certainly not a major flaw. If it is the row's only flag reason, the severity is informational (minor at most). If it co-occurs with another flagged axis, judge severity from that other axis alone, as if axis 6 were ideal -- never let CLEAR\_COMMIT\_TEXT push a row up a tier.

\emph{What to do}

For each file: identify which axis or axes are non-ideal (that's why flag = yes), then apply the tier rules above using only what the row's own notes already say about verification depth. For every row, report: the test's identity (copied verbatim from its test\_id column); which axis or axes drove the flag (e.g.~``axis2 = MODERATE''); the severity tier; whether the proof behind it was live-verified, source-only, or not applicable; and one to three sentences of justification that cite the specific evidence already present in the row's notes -- don't derive new evidence, just cite what's there.
\end{minipage}}
\caption{Instructions for the Step 1 (proposal) severity scoring agent.}\label{fig:qualitative-severity1-instructions}
\end{figure*}

\begin{figure*}[tb]
\footnotesize
\noindent\fbox{\begin{minipage}{\dimexpr\linewidth-2\fboxsep-2\fboxrule\relax}
You are one of three independent panelists re-deriving whether a specific test deserves a critical severity tier for its test-quality defect. You are blind to any prior proposal or reasoning about this row -- form your own judgment from scratch by reading the actual test and the production code it exercises yourself.

Row identity you've been given:

Test identity: {[}test\_id{]}
File: {[}file path{]}
Commit: {[}commit hash{]}
Axis or axes flagged by the original scoring (this tells you what to look at, but is not itself a severity claim): {[}flagged axes{]}
Reading the code: read the test's source and the production code it exercises, at this exact commit, read-only, from this corpus's own checkout.

Verifying live, if you need to: if you want to check a claim empirically, create your own dedicated, throwaway worktree at this row's own commit. Run the real test against your worktree using this corpus's own established test-running procedure (its particular Python version, environment variables, and any test-runner wrapper it requires -- follow this corpus's own setup exactly, not a generic pytest invocation, since some corpora need a specific harness to get a correct result).

The bar for critical -- narrow, and only these three patterns qualify:

\begin{itemize}
\tightlist
\item
  The test's assertion checks something unrelated to its stated claim (a mismatch), and you can show this concretely -- either by running it live, or by tracing precisely what the assertion can and cannot distinguish.
\item
  The test's setup constructs a fundamentally different scenario than the one it claims to exercise (not just excess setup), and a live check proves it -- for example, patching the real logic the test claims to cover doesn't make the test fail, because its setup never actually built that scenario in the first place.
\item
  You can hand-patch the specific logic the assertion is supposed to be checking, rerun the real test, and it still passes -- meaning you personally reproduce a live failure-to-catch, not cite someone else's claim of one.
\end{itemize}

Your vote: vote critical only if you personally verified one of these three patterns, live, yourself, right now -- not because the row's axis verdicts merely look concerning on paper. If you cannot reproduce a live failure-to-catch within reasonable effort, vote not critical. Report your verdict together with a one-paragraph justification explaining exactly what you did and what you found.
\end{minipage}}
\caption{Instructions for the Step 2 (three panelists) severity score review agents.}\label{fig:qualitative-severity2-instructions}
\end{figure*}

\begin{table*}[tb]
\footnotesize
\centering
\caption{Changes in qualitative rubric scoring from Stage A to Stage B.}\label{tab:stage-ab-changes}
\begin{tabular}{@{}
  >{\raggedright\arraybackslash}p{(\linewidth - 6\tabcolsep) * \real{0.1859}}
  >{\raggedright\arraybackslash}p{(\linewidth - 6\tabcolsep) * \real{0.2180}}
  >{\raggedright\arraybackslash}p{(\linewidth - 6\tabcolsep) * \real{0.2546}}
  >{\raggedright\arraybackslash}p{(\linewidth - 6\tabcolsep) * \real{0.3415}}
@{}}
\toprule
\begin{minipage}[b]{\linewidth}\raggedright
Severity
\end{minipage} & \begin{minipage}[b]{\linewidth}\raggedright
Total tests
\end{minipage} & \begin{minipage}[b]{\linewidth}\raggedright
Unchanged (A=B)
\end{minipage} & \begin{minipage}[b]{\linewidth}\raggedright
Changed (\(\geq 1\) axis downgraded)
\end{minipage} \\
\midrule
\multicolumn{4}{>{\raggedright\arraybackslash}p{\linewidth-2\tabcolsep}}{\textbf{Extractor}}\\
critical & 6 & 4 & 2 \\
moderate & 58 & 52 & 6 \\
\midrule
\multicolumn{4}{>{\raggedright\arraybackslash}p{\linewidth-2\tabcolsep}}{\textbf{Django}}\\
critical & 5 & 3 & 2 \\
moderate & 44 & 37 & 7 \\
\bottomrule
\end{tabular}
\normalsize
\end{table*}

\begin{table*}[tb]
\footnotesize
\centering
\caption{Breakdown by rubric axis of changes in scoring from Stage A to Stage B.}\label{tab:stage-ab-changes2}
\begin{tabular}{@{}
  >{\raggedright\arraybackslash}p{(\linewidth - 6\tabcolsep) * \real{0.1637}}
  >{\raggedright\arraybackslash}p{(\linewidth - 6\tabcolsep) * \real{0.2951}}
  >{\raggedright\arraybackslash}p{(\linewidth - 6\tabcolsep) * \real{0.1531}}
  >{\raggedright\arraybackslash}p{(\linewidth - 6\tabcolsep) * \real{0.3882}}
@{}}
\toprule
\begin{minipage}[b]{\linewidth}\raggedright
Severity
\end{minipage} & \begin{minipage}[b]{\linewidth}\raggedright
Axis changed
\end{minipage} & \begin{minipage}[b]{\linewidth}\raggedright
\# tests
\end{minipage} & \begin{minipage}[b]{\linewidth}\raggedright
Direction
\end{minipage} \\
\midrule
\multicolumn{4}{>{\raggedright\arraybackslash}p{\linewidth-2\tabcolsep}}{\textbf{Extractor}}\\
critical & axis2 (assertion strength) & 2 & STRICT -\textgreater{} WEAK / MODERATE \\
moderate & axis2 (assertion strength) & 5 & STRICT -\textgreater{} MODERATE (4×) / WEAK (1×) \\
moderate & axis5 (scenario realism) & 1 & OK -\textgreater{} DEFECTIVE \\
\midrule
\multicolumn{4}{>{\raggedright\arraybackslash}p{\linewidth-2\tabcolsep}}{\textbf{Django}}\\
critical & axis5 (scenario realism) & 2 & OK -\textgreater{} DEFECTIVE \\
moderate & axis2 (assertion strength) & 6 & STRICT -\textgreater{} MODERATE \\
moderate & axis5 (scenario realism) & 1 & OK -\textgreater{} DEFECTIVE \\
\bottomrule
\end{tabular}
\normalsize
\end{table*}

\subsection{Historic-revert vs.~qualitative-rubric detail tables}\label{sec:test-details}

\subsubsection{Extractor}\label{extractor}

See Tables \ref{tab:extractor-weak-axis} and \ref{tab:extractor-non-effective}.

\begin{table*}[tb]
\footnotesize
\centering
\caption{Extractor tests flagged assertion-strength WEAK by the qualitative rubric, with historical-revert outcomes. Historic-revert: S = SURVIVED, E = EFFECTIVE, E/New = EFFECTIVE (TESTING\_NEW\_FUNCTION).}\label{tab:extractor-weak-axis}
\begin{tabular}{@{}
  >{\raggedright\arraybackslash}p{(\linewidth - 6\tabcolsep) * \real{0.2271}}
  >{\raggedright\arraybackslash}p{(\linewidth - 6\tabcolsep) * \real{0.1155}}
  >{\raggedright\arraybackslash}p{(\linewidth - 6\tabcolsep) * \real{0.0986}}
  >{\raggedright\arraybackslash}p{(\linewidth - 6\tabcolsep) * \real{0.5588}}
@{}}
\toprule
\begin{minipage}[b]{\linewidth}\raggedright
Test (commit)
\end{minipage} & \begin{minipage}[b]{\linewidth}\raggedright
Historic-revert
\end{minipage} & \begin{minipage}[b]{\linewidth}\raggedright
Severity
\end{minipage} & \begin{minipage}[b]{\linewidth}\raggedright
Commentary
\end{minipage} \\
\midrule
\multicolumn{4}{>{\raggedright\arraybackslash}p{\linewidth-2\tabcolsep}}{\textbf{Weak-value/loose-bound oracle}}\\
\texttt{\seqsplit{test\_flags\_struct\_behind\_name\_still\_detected}} & \seqsplit{S} & \seqsplit{moderate} & There is one assertion, and the pre-commit (loose) and post-commit casing rules both return \texttt{\seqsplit{True}} for the input used by the test.  \\
\texttt{\seqsplit{test\_forward\_branch\_handler\_dominance}} & \seqsplit{E/New} & \seqsplit{critical} & Both test assertions are structurally non-discriminating. \\
\texttt{\seqsplit{test\_by\_file\_breaks\_the\_package\_merge\_cycle}} & \seqsplit{E} & \seqsplit{moderate} & Test assertions only check import-set membership on two files and never check for all 3 expected filenames. EFFECTIVE only because reverting the commit is a mutation that these assertions happen to catch. \\
\texttt{\seqsplit{test\_qualified\_keys\_avoid\_short\_name\_collision}} & \seqsplit{E} & \seqsplit{moderate} & Reverting the commit re-merges three types into one, which the test's count check catches, hence EFFECTIVE. But the test never runs the other half of the fix, the code that fills in field data without guessing between same-named types; breaking that half goes unnoticed. \\
\midrule
\multicolumn{4}{>{\raggedright\arraybackslash}p{\linewidth-2\tabcolsep}}{\textbf{Wrong entry point / bypassed wrapper}}\\
\texttt{\seqsplit{test\_suffix\_must\_be\_a\_real\_suffix}} & \seqsplit{E/New} & \seqsplit{critical} & Test only checks fields with no empty-string key in \texttt{\seqsplit{\_ENC}}, so an earlier dict-miss fallback already returns \texttt{\seqsplit{False}} before the new length guard is ever reached. \\
\texttt{\seqsplit{test\_word\_substitution}} & \seqsplit{E/New} & \seqsplit{moderate} & The decoder has two branches (lowercase continuation, uppercase terminator); the test routes only through the uppercase branch, the lowercase route is never entered. \\
\texttt{\seqsplit{test\_lendelim\_tag\_wire\_type\_2\_is\_plain}} & \seqsplit{S} & \seqsplit{moderate} & The production code falls back to a ``plain'' default whenever none of its more specific detection rules match. The test's input is a length-delimited field, a case none of those specific rules are meant to fire on anyway, so it gets the ``plain'' default regardless of whether the surrounding detection logic is present, working, or missing entirely. \\
\bottomrule
\end{tabular}
\normalsize
\end{table*}

\begin{table*}[tb]
\footnotesize
\centering
\caption{Extractor tests flagged non-EFFECTIVE by historical-revert, with qualitative rubric scores. Historic-revert: S = SURVIVED, U = UNREACHED.}\label{tab:extractor-non-effective}
\begin{tabular}{@{}
  >{\raggedright\arraybackslash}p{(\linewidth - 6\tabcolsep) * \real{0.2630}}
  >{\raggedright\arraybackslash}p{(\linewidth - 6\tabcolsep) * \real{0.1184}}
  >{\raggedright\arraybackslash}p{(\linewidth - 6\tabcolsep) * \real{0.0988}}
  >{\raggedright\arraybackslash}p{(\linewidth - 6\tabcolsep) * \real{0.5198}}
@{}}
\toprule
\begin{minipage}[b]{\linewidth}\raggedright
Test (commit)
\end{minipage} & \begin{minipage}[b]{\linewidth}\raggedright
Historic-revert
\end{minipage} & \begin{minipage}[b]{\linewidth}\raggedright
Severity
\end{minipage} & \begin{minipage}[b]{\linewidth}\raggedright
Commentary
\end{minipage} \\
\midrule
\multicolumn{4}{>{\raggedright\arraybackslash}p{\linewidth-2\tabcolsep}}{\textbf{Weak-value/loose-bound oracle}}\\
\texttt{\seqsplit{test\_flags\_struct\_behind\_name\_still\_detected}} & \seqsplit{S} & \seqsplit{moderate} & See Table \ref{tab:extractor-weak-axis}. \\
\texttt{\seqsplit{test\_varint\_tag\_shift\_is\_plain}} & \seqsplit{S} & -- & Test claims to check one specific safeguard, but its input also differs from the bad case in a second, unrelated way. That second difference is enough to make the test pass, whether or not the safeguard works. \\
\texttt{\seqsplit{test\_index\_scale\_lsl3\_on\_nonfield\_reg\_does\_not\_block\_fixed}} & \seqsplit{S} & \seqsplit{moderate} & The test's one assertion (a decoy index-scale \texttt{\seqsplit{lsl}} doesn't block fixed32 classification) passes equally whether the terminator's register-matching is a genuine, selective gate or simply absent. \\
\texttt{\seqsplit{test\_lendelim\_tag\_wire\_type\_2\_is\_plain}} & \seqsplit{S} & \seqsplit{moderate} & See Table \ref{tab:extractor-weak-axis}. \\
\midrule
\multicolumn{4}{>{\raggedright\arraybackslash}p{\linewidth-2\tabcolsep}}{\textbf{No defect}}\\
\texttt{\seqsplit{test\_xasstring\_routed\_field\_not\_rebiased}} & \seqsplit{S} & -- & Test checks that one kind of field is left unchanged by a new feature. The feature is written to skip fields of that kind. Reverting the commit removes the feature, skip-check included, so this field is unchanged either way. The test is well written yet SURVIVED is also a correct verdict. \\
\texttt{\seqsplit{test\_no\_package\_note\_when\_no\_package}} & \seqsplit{U} & -- & The test is well written and checks what it claims. The commit adds a check which is not exercised by the test so SURVIVED is also a correct verdict. \\
\texttt{\seqsplit{test\_no\_recon\_caveat\_when\_no\_enums}} & \seqsplit{S} & -- & Structurally the same as the row above. \\
\texttt{\seqsplit{test\_explicit\_prefixes\_override\_default}} & \seqsplit{S} & -- & A value could be passed as an argument or set globally. The commit deletes the global path; this test only exercises the argument path. The test is well written yet SURVIVED is also a correct verdict. \\
\bottomrule
\end{tabular}
\normalsize
\end{table*}

\subsubsection{Django}\label{django}

See Tables \ref{tab:django-weak-axis} and \ref{tab:django-non-effective}.

\begin{table*}[tb]
\footnotesize
\centering
\caption{Django tests flagged assertion-strength WEAK by the qualitative rubric, with historical-revert outcomes. Historic-revert: S = SURVIVED, E = EFFECTIVE.}\label{tab:django-weak-axis}
\begin{tabular}{@{}
  >{\raggedright\arraybackslash}p{(\linewidth - 6\tabcolsep) * \real{0.2121}}
  >{\raggedright\arraybackslash}p{(\linewidth - 6\tabcolsep) * \real{0.1011}}
  >{\raggedright\arraybackslash}p{(\linewidth - 6\tabcolsep) * \real{0.0976}}
  >{\raggedright\arraybackslash}p{(\linewidth - 6\tabcolsep) * \real{0.5891}}
@{}}
\toprule
\begin{minipage}[b]{\linewidth}\raggedright
Test (commit)
\end{minipage} & \begin{minipage}[b]{\linewidth}\raggedright
Historic-revert
\end{minipage} & \begin{minipage}[b]{\linewidth}\raggedright
Severity
\end{minipage} & \begin{minipage}[b]{\linewidth}\raggedright
Commentary
\end{minipage} \\
\midrule
\multicolumn{4}{>{\raggedright\arraybackslash}p{\linewidth-2\tabcolsep}}{\textbf{Weak-value/loose-bound oracle}}\\
\texttt{\seqsplit{test\_change\_password}} & \seqsplit{E} & \seqsplit{critical} & The test only checks \texttt{\seqsplit{assertIsInstance(user,\ User)}} after a password change, but it never actually changes the password, so the session hash it is meant to validate never diverges regardless of whether the fix's \texttt{\seqsplit{await\ request.auser()}} is present or reverted to sync \texttt{\seqsplit{request.user}}. EFFECTIVE here is coincidental: reverting the whole commit instead removes the \texttt{\seqsplit{auser()}} call entirely, crashing with \texttt{\seqsplit{AttributeError}} before the weak assertion (or the narrow oracle gap) is ever reached. \\
\texttt{\seqsplit{test\_aggregate\_subquery\_annotation}} & \seqsplit{S} & \seqsplit{critical} & \texttt{assertLessEqual(sql.count(\textquotesingle{}SELECT\textquotesingle{}),\ 4)} accepts both the buggy count (4) and the fixed count (3), so it cannot catch the bug it was written for. \\
\texttt{\seqsplit{test\_remove\_unique\_together\_on\_pk\_field}} & \seqsplit{S} & \seqsplit{critical} & Fully reverting the commit's fix, removing the exact behaviour the test is meant to guard, still leaves the test passing. \\
\texttt{\seqsplit{test\_rename\_model\_with\_db\_table\_rename\_m2m}} & \seqsplit{E} & \seqsplit{moderate} & The test only checks that the rename completes without raising. A concrete, equally simple check is available and unused: asserting the actual row count on the renamed relation, which would additionally catch corruption that doesn't raise. \\
\texttt{\seqsplit{test\_namespaced\_db\_table\_foreign\_key\_reference}} & \seqsplit{E} & \seqsplit{moderate} & The test has no explicit assertions, it passes as long as the model/field creation calls complete without raising. \\
\texttt{\seqsplit{test\_alter\_field\_with\_custom\_db\_type}} & \seqsplit{E} & \seqsplit{moderate} & The test has no explicit assertions, it passes as long as the field alteration completes without raising. It never checks that the resulting column actually reflects the new definition. \\
\texttt{\seqsplit{test\_server\_formatter\_default\_format}} & \seqsplit{E} & \seqsplit{moderate} & One of the test's two assertions checks the exact formatted log line; the other, covering the date format this commit actually changes, only checks a broad regex that would match almost any bracketed date-like string, not the specific format the commit introduces. \\
\texttt{\seqsplit{test\_rename\_m2m\_field\_with\_2\_references}} & \seqsplit{E} & \seqsplit{moderate} & The test does correctly catch corruption on the rename's crash-path, but that's not what its gap is about: after the rename, it never checks that foreign-key targets and data actually survived, only that the rename itself didn't crash. \\
\texttt{\seqsplit{test\_parameter\_count\_exceeds\_variable\_or\_column\_limit}} & \seqsplit{S} & \seqsplit{moderate} & The only check is implicit, the test passes as long as the call doesn't raise, and the return value itself is never inspected. \\
\bottomrule
\end{tabular}
\normalsize
\end{table*}

\begin{table*}[tb]
\scriptsize
\centering
\caption{Django tests flagged non-EFFECTIVE by historical-revert, with qualitative rubric scores. Historic-revert: S = SURVIVED, U = UNREACHABLE.}\label{tab:django-non-effective}
\begin{tabular}{@{}
  >{\raggedright\arraybackslash}p{(\linewidth - 6\tabcolsep) * \real{0.1895}}
  >{\raggedright\arraybackslash}p{(\linewidth - 6\tabcolsep) * \real{0.0964}}
  >{\raggedright\arraybackslash}p{(\linewidth - 6\tabcolsep) * \real{0.0964}}
  >{\raggedright\arraybackslash}p{(\linewidth - 6\tabcolsep) * \real{0.6177}}
@{}}
\toprule
\begin{minipage}[b]{\linewidth}\raggedright
Test (commit)
\end{minipage} & \begin{minipage}[b]{\linewidth}\raggedright
Historic-revert
\end{minipage} & \begin{minipage}[b]{\linewidth}\raggedright
Severity
\end{minipage} & \begin{minipage}[b]{\linewidth}\raggedright
Commentary
\end{minipage} \\
\midrule
\multicolumn{4}{>{\raggedright\arraybackslash}p{\linewidth-2\tabcolsep}}{\textbf{Dead assertion order (stacked context managers)}}\\
\texttt{\seqsplit{test\_json\_response\_raises\_type\_error\_with\_safe\_arg}} & \seqsplit{S} & \seqsplit{critical} & The test wraps two checks in one \texttt{\seqsplit{with\ (assertRaisesMessage(...),\ assertWarnsMessage(...)):}} block. Python exits stacked context managers in reverse order, so the warning check (entered second) exits first, sees the exception still active, and skips its own check; the warning half can never fail. The other half also cannot distinguish old from new code: it passes \texttt{\seqsplit{safe=True}} explicitly, and both versions raise the same error for that regardless of the fix. \\
\midrule
\multicolumn{4}{>{\raggedright\arraybackslash}p{\linewidth-2\tabcolsep}}{\textbf{Weak-value/loose-bound oracle}}\\
\texttt{\seqsplit{test\_parameter\_count\_exceeds\_variable\_or\_column\_limit}} & \seqsplit{S} & \seqsplit{moderate} & See Table \ref{tab:django-weak-axis}. \\
\texttt{\seqsplit{test\_aggregate\_subquery\_annotation}} & \seqsplit{S} & \seqsplit{critical} & Same test as the WEAK-axis table above. \\
\texttt{\seqsplit{test\_remove\_unique\_together\_on\_pk\_field}} & \seqsplit{S} & \seqsplit{critical} & Same test as the WEAK-axis table above. \\
\texttt{\seqsplit{test\_db\_default\_equivalent\_sql\_noop}} & \seqsplit{S} & \seqsplit{critical} & \texttt{\seqsplit{db\_default}} is not a \texttt{\seqsplit{non\_db\_attr}}, so the new guard's condition genuinely evaluates and its \texttt{\seqsplit{pop()}} calls genuinely execute here. But \texttt{\seqsplit{Field.\_\_init\_\_}} already normalises a bare string \texttt{\seqsplit{db\_default="foo"}} into \texttt{\seqsplit{Value("foo")}} before \texttt{\seqsplit{deconstruct()}} ever runs, so \texttt{old\_kwargs{[}"db\_default"{]}} and \texttt{new\_kwargs{[}"db\_default"{]}} are already equal Python objects whether or not this commit's \texttt{\seqsplit{pop()}} calls fire; popping two already-equal values changes nothing. \texttt{\seqsplit{assertNumQueries(0)}} is precise, but this fixture cannot tell the new code's effect apart from its absence -- live-verified by hand-patching the fix's comparison and rerunning: still zero queries. \\
\midrule
\multicolumn{4}{>{\raggedright\arraybackslash}p{\linewidth-2\tabcolsep}}{\textbf{No defect}}\\
\texttt{\seqsplit{test\_check\_constraints}} & \seqsplit{S} & -- & The fix changes whether Django loops over all configured databases or just the ones under test when checking constraints; this only changes the warning count on backends that lack table-check-constraint support. When run with SQLite, which supports them, the expected result is identical either way. A backend-coverage gap in the test environment, not a flaw in the test. \\
\texttt{\seqsplit{test\_nested\_key\_transform\_on\_subquery}} & \seqsplit{U} & -- & The commit's one changed line sits on a code path this test's query never reaches, so the test can't see it either way. \\
\texttt{\seqsplit{test\_check\_database\_version\_supported}} & \seqsplit{U} & -- & The test's own code path never reaches the line this commit changes, so it can't see the difference either way. \\
\texttt{\seqsplit{test\_composite\_primary\_key\_not\_unique\_together}} & \seqsplit{U} & -- & The commit's one changed line sits on a code path this test's query never reaches, so the test can't see it either way. \\
\texttt{\seqsplit{test\_in}} & \seqsplit{S} & -- & The commit's change lives entirely in Oracle-specific code for translating JSON lookups; this test runs against SQLite, which never calls that code at all, so nothing the commit touches is exercised. \\
\texttt{\seqsplit{test\_order\_by\_case\_when\_constant\_value}} & \seqsplit{S} & -- & The commit reorganizes how the SQL compiler builds the fallback value for an ordering expression, but computes exactly the same value both before and after -- it's an internal rewrite, not a behaviour change. The test correctly checks that value, so it passes either way. \\
\texttt{\seqsplit{test\_context\_copyable}} & \seqsplit{S} & -- & The commit swaps a manual rebuild of the copied object's class and attributes for a cleaner, equivalent call to the parent class's own copy method. Both produce the same copy on the Python version this test runs under, so the test can't see any difference. \\
\texttt{\seqsplit{test\_union\_all\_none\_slice}} & \seqsplit{S} & -- & The commit changes how an empty half of a union query is built internally, but an empty half contributes no rows to the result either way. The test checks the returned rows, not how the query was built, so it passes regardless. \\
\texttt{\seqsplit{test\_union\_none\_slice}} & \seqsplit{S} & -- & Same shape as the row above: the commit changes how an empty half of a union is built internally, but empty means zero rows either way. The test checks the returned rows, not the internal SQL, so it can't tell the two versions apart. \\
\texttt{\seqsplit{test\_order\_by\_expression\_ref}} & \seqsplit{S} & -- & The commit changes how \texttt{\seqsplit{order\_by}} references are resolved, but only for a specific, ambiguous case this test doesn't set up. For the plain case this test does check, the old and new logic produce the same ordering, so the test passes either way. \\
\bottomrule
\end{tabular}
\normalsize
\end{table*}

\subsubsection{Pandas}\label{pandas}

See Tables \ref{tab:pandas-weak-axis} and \ref{tab:pandas-non-effective}.

\begin{table*}[tb]
\footnotesize
\centering
\caption{Pandas tests flagged assertion-strength WEAK by the qualitative rubric, with historical-revert outcome. Historic-revert: E = EFFECTIVE.}\label{tab:pandas-weak-axis}
\begin{tabular}{@{}
  >{\raggedright\arraybackslash}p{(\linewidth - 6\tabcolsep) * \real{0.2310}}
  >{\raggedright\arraybackslash}p{(\linewidth - 6\tabcolsep) * \real{0.1196}}
  >{\raggedright\arraybackslash}p{(\linewidth - 6\tabcolsep) * \real{0.0988}}
  >{\raggedright\arraybackslash}p{(\linewidth - 6\tabcolsep) * \real{0.5505}}
@{}}
\toprule
\begin{minipage}[b]{\linewidth}\raggedright
Test (commit)
\end{minipage} & \begin{minipage}[b]{\linewidth}\raggedright
Historic-revert
\end{minipage} & \begin{minipage}[b]{\linewidth}\raggedright
Severity
\end{minipage} & \begin{minipage}[b]{\linewidth}\raggedright
Commentary
\end{minipage} \\
\midrule
\multicolumn{4}{>{\raggedright\arraybackslash}p{\linewidth-2\tabcolsep}}{\textbf{No assertion beyond non-crash}}\\
\texttt{\seqsplit{test\_pass\_colormap\_instance}} & \seqsplit{E} & \seqsplit{moderate} & Calls \texttt{\seqsplit{df.plot.scatter}}/\texttt{\seqsplit{hexbin}} with a \texttt{\seqsplit{Colormap}} \emph{instance} (rather than a string name) and checks nothing beyond the call not raising. A concrete stricter check exists and is mechanically straightforward: \texttt{ax.collections{[}-1{]}.get\_cmap()\ is\ cmap} on the returned \texttt{\seqsplit{Axes}}. \\
\texttt{\seqsplit{test\_interchange\_from\_corrected\_buffer\_dtypes}} & \seqsplit{E} & \seqsplit{moderate} & Monkeypatches a string column's reported buffer dtype to simulate a corrected interchange scenario, then calls \texttt{\seqsplit{pd.api.interchange.from\_dataframe(df)}} without even assigning the return value: the only signal is that the call doesn't raise. \\
\texttt{\seqsplit{test\_xz\_compression\_level\_read}} & \seqsplit{E} & \seqsplit{moderate} & The size-comparison assertion (\texttt{compressed\_size\_default\ \textless{}\ compressed\_size\_fast}) is well-calibrated, but the \texttt{\seqsplit{to\_csv}} branch's read-back call, \texttt{\seqsplit{pd.read\_csv(path,\ compression="xz")}}, is never checked against the original data: unlike the file's sibling round-trip test, which does assert equality. \\
\bottomrule
\end{tabular}
\normalsize
\end{table*}

\begin{table*}[tb]
\footnotesize
\centering
\caption{Pandas tests flagged non-EFFECTIVE by historical-revert, with qualitative rubric scores. Historic-revert: S = SURVIVED, U = UNREACHED.}\label{tab:pandas-non-effective}
\begin{tabular}{@{}
  >{\raggedright\arraybackslash}p{(\linewidth - 6\tabcolsep) * \real{0.2209}}
  >{\raggedright\arraybackslash}p{(\linewidth - 6\tabcolsep) * \real{0.1078}}
  >{\raggedright\arraybackslash}p{(\linewidth - 6\tabcolsep) * \real{0.1003}}
  >{\raggedright\arraybackslash}p{(\linewidth - 6\tabcolsep) * \real{0.5710}}
@{}}
\toprule
\begin{minipage}[b]{\linewidth}\raggedright
Test (commit)
\end{minipage} & \begin{minipage}[b]{\linewidth}\raggedright
Historic-revert
\end{minipage} & \begin{minipage}[b]{\linewidth}\raggedright
Severity
\end{minipage} & \begin{minipage}[b]{\linewidth}\raggedright
Commentary
\end{minipage} \\
\midrule
\multicolumn{4}{>{\raggedright\arraybackslash}p{\linewidth-2\tabcolsep}}{\textbf{Wrong entry point / bypassed wrapper}}\\
\texttt{\seqsplit{test\_fillna\_dict\_inplace\_nonunique\_columns}} & \seqsplit{S} & \seqsplit{moderate} & The test's three duplicate \texttt{\seqsplit{"A"}} columns sit adjacent, so \texttt{\seqsplit{df.columns.get\_loc("A")}} returns a plain contiguous slice. For a contiguous slice, the old single-write and the new per-location-loop write touch the same positions in the same order and produce identical output. The fix's real target, non-contiguous duplicate columns, is never constructed here. \\
\texttt{\seqsplit{test\_astype\_numpy\_to\_ea}} & \seqsplit{U} & \seqsplit{informational} & For this test's direction (numpy int64 -\textgreater{} \texttt{\seqsplit{Int64Dtype}}), the code finds a value before the buggy fallback line runs, so that line is never reached. The typo this commit fixes only manifests in the reverse direction, which this test does not exercise. \\
\texttt{\seqsplit{test\_dt\_round\_nonnano\_higher\_resolution\_no\_op}} & \seqsplit{S} & \seqsplit{informational} & The fix is a one-line change inside the array-level \texttt{\seqsplit{\_round()}}, and the test's \texttt{\seqsplit{assert\ not\ np.shares\_memory(...)}} looks like a precise oracle for exactly that line, but it observes memory identity only after the result passes back up through the \texttt{\seqsplit{.dt}} accessor's Series-construction wrapper, one layer above the fixed line, which already produces a fresh array regardless of what the inner method returns. \\
\texttt{\seqsplit{test\_duration\_fillna\_numpy}} & \seqsplit{S} & -- & The test fills with a whole \texttt{\seqsplit{TimedeltaArray}}, not a scalar \texttt{\seqsplit{Timedelta}}. The commit's refactor only changes behaviour for a scalar-\texttt{\seqsplit{Timedelta}}-with-unit input; the array-shaped branch this test exercises is identical in both old and new code. \\
\texttt{\seqsplit{test\_astype\_dt64\_to\_int64}} & \seqsplit{S} & -- & \texttt{M8{[}ns{]}\ -\textgreater{}\ int64} is a lossless direct cast under both the removed and the new code path for this test's dtype pair, including the non-NaT-fill-value case this commit adds. The original bug was that old code \emph{raised} for this dtype combination, not that it computed a wrong value; this test's pairing does not trigger the raise either way. \\
\midrule
\multicolumn{4}{>{\raggedright\arraybackslash}p{\linewidth-2\tabcolsep}}{\textbf{Weak-value/loose-bound oracle}}\\
\texttt{\seqsplit{test\_private\_values\_dt64\_multiblock}} & \seqsplit{S} & \seqsplit{informational} & The fix's whole point is that the DataFrame should still have 2 separate internal blocks \emph{after} \texttt{\seqsplit{df.\_values}} is accessed. The test checks the block count only \emph{before} the call, then checks only the returned values afterwards: old code's silent re-consolidation does not change the decoded values, so nothing catches the regression the commit targets. \\
\texttt{\seqsplit{test\_stack\_order\_with\_unsorted\_levels\_multi\_row\_2}} & \seqsplit{S} & -- & Reproduced both code paths directly: for this test's 2-distinct-value column level, the old and new internal code/level pairings decode to the exact same per-row values; only the internal array orientation differs, which the test's frame-equality check never inspects. \\
\bottomrule
\end{tabular}
\normalsize
\end{table*}

\subsection{Low-kill-rate triage detail tables}\label{sec:lowkillrate-detail}

Tables \ref{tab:percommit-triage} and \ref{tab:covguided-triage} give the Extractor tests found defective by low-kill-rate triage (\S\ref{ast-mutation-scoring}), under the per-commit and coverage-guided mutation evaluations respectively; two of the five (marked \textsuperscript{†}) independently confirm defects already found by the other evaluations rather than being wholly new.

\begin{table*}[tb]
\scriptsize
\centering
\caption{Extractor tests found defective by per-commit low-kill-rate triage. Historic-revert: S = SURVIVED, E/New = EFFECTIVE (TESTING\_NEW\_FUNCTION).}\label{tab:percommit-triage}
\begin{tabular}{@{}
  >{\raggedright\arraybackslash}p{(\linewidth - 10\tabcolsep) * \real{0.1720}}
  >{\raggedright\arraybackslash}p{(\linewidth - 10\tabcolsep) * \real{0.0942}}
  >{\raggedleft\arraybackslash}p{(\linewidth - 10\tabcolsep) * \real{0.0942}}
  >{\raggedright\arraybackslash}p{(\linewidth - 10\tabcolsep) * \real{0.0942}}
  >{\raggedright\arraybackslash}p{(\linewidth - 10\tabcolsep) * \real{0.1322}}
  >{\raggedright\arraybackslash}p{(\linewidth - 10\tabcolsep) * \real{0.4132}}
@{}}
\toprule
\begin{minipage}[b]{\linewidth}\raggedright
Test (commit)
\end{minipage} & \begin{minipage}[b]{\linewidth}\raggedright
Historic-revert
\end{minipage} & \begin{minipage}[b]{\linewidth}\raggedleft
Kill rate
\end{minipage} & \begin{minipage}[b]{\linewidth}\raggedright
A2 Strength
\end{minipage} & \begin{minipage}[b]{\linewidth}\raggedright
Category
\end{minipage} & \begin{minipage}[b]{\linewidth}\raggedright
Commentary
\end{minipage} \\
\midrule
\texttt{\seqsplit{test\_varint\_tag\_shift\_is\_plain}}\textsuperscript{†} & S & 0/14 & STRICT & Weak-value/loose-bound oracle & The test's input differs from the bad case in a second, unrelated way that alone makes it pass, masking whether the actual safeguard works. \\
\texttt{\seqsplit{test\_no\_sibling\_data\_ivar\_is\_not\_phantom}} & \seqsplit{E/New} & 0/4 & STRICT & Weak-value/loose-bound oracle & The fixture's dict lookup misses regardless of how the surrounding guards are perturbed, so the negative assertion cannot discriminate correct from broken. \\
\texttt{\seqsplit{test\_real\_vmaddr\_is\_not\_a\_bind}} & \seqsplit{E/New} & 0/11 & STRICT & Weak-value/loose-bound oracle & The fixture value sits far from the \texttt{1\textless{}\textless{}62} boundary the guard exists to test, so the guard's own comparison is never exercised at a discriminating point. \\
\texttt{\seqsplit{test\_resolve\_image\_exact\_full\_path}}\textsuperscript{†} & \seqsplit{E/New} & 0/8 & MODERATE & Wrong entry point / bypassed wrapper & The single-image fixture only ever exercises the exact-match tier; it cannot tell whether the other two resolution tiers work at all. \\
\texttt{\seqsplit{test\_oracle\_contradiction\_rejected}} & \seqsplit{E/New} & 0/21 & STRICT & Wrong entry point / bypassed wrapper & The code recovers a message's field numbers, then double-checks them against a second, independent count of the same numbers, rejecting if they do not line up. The test's example has the two counts share nothing in common at all, such an extreme mismatch that almost any check would catch it, so passing does not prove the actual comparison is correct. \\
\bottomrule
\end{tabular}
\normalsize
\end{table*}

\textsuperscript{†}The qualitative evaluation flags \texttt{\seqsplit{test\_resolve\_image\_exact\_full\_path}} as the same defect, and as having critical severity (\texttt{\seqsplit{MODERATE}} assertion strength but \texttt{\seqsplit{axis1=PARTIAL}} and \texttt{\seqsplit{axis5=DEFECTIVE}}). \texttt{\seqsplit{test\_varint\_tag\_shift\_is\_plain}} is flagged by the historic-revert evaluation.

\begin{table*}[tb]
\scriptsize
\centering
\caption{Extractor tests found defective by coverage-guided low-kill-rate triage. Historic-revert: S = SURVIVED, E/New = EFFECTIVE (TESTING\_NEW\_FUNCTION).}\label{tab:covguided-triage}
\begin{tabular}{@{}
  >{\raggedright\arraybackslash}p{(\linewidth - 10\tabcolsep) * \real{0.2650}}
  >{\raggedright\arraybackslash}p{(\linewidth - 10\tabcolsep) * \real{0.1314}}
  >{\raggedleft\arraybackslash}p{(\linewidth - 10\tabcolsep) * \real{0.1018}}
  >{\raggedright\arraybackslash}p{(\linewidth - 10\tabcolsep) * \real{0.1125}}
  >{\raggedright\arraybackslash}p{(\linewidth - 10\tabcolsep) * \real{0.2035}}
  >{\raggedright\arraybackslash}p{(\linewidth - 10\tabcolsep) * \real{0.1858}}
@{}}
\toprule
\begin{minipage}[b]{\linewidth}\raggedright
Test (commit)
\end{minipage} & \begin{minipage}[b]{\linewidth}\raggedright
Historic-revert
\end{minipage} & \begin{minipage}[b]{\linewidth}\raggedleft
Kill rate
\end{minipage} & \begin{minipage}[b]{\linewidth}\raggedright
A2 Strength
\end{minipage} & \begin{minipage}[b]{\linewidth}\raggedright
Category
\end{minipage} & \begin{minipage}[b]{\linewidth}\raggedright
Commentary
\end{minipage} \\
\midrule
\texttt{\seqsplit{test\_varint\_tag\_shift\_is\_plain}}\textsuperscript{†} & S & 0/38 & STRICT & Weak-value/loose-bound oracle & Same defect as in table above \\
\texttt{\seqsplit{test\_no\_sibling\_data\_ivar\_is\_not\_phantom}} & \seqsplit{E/New} & 0/4 & STRICT & Weak-value/loose-bound oracle & Same defect as in table above. \\
\texttt{\seqsplit{test\_real\_vmaddr\_is\_not\_a\_bind}} & \seqsplit{E/New} & 0/3 & STRICT & Weak-value/loose-bound oracle & Same defect as in table above. \\
\texttt{\seqsplit{test\_resolve\_image\_exact\_full\_path}}\textsuperscript{†} & \seqsplit{E/New} & 12/40 & MODERATE & Wrong entry point / bypassed wrapper & Same defect as in table above. \\
\bottomrule
\end{tabular}
\normalsize
\end{table*}

\subsection{Low-kill-rate defect-mining agent instructions}\label{low-kill-rate-defect-mining-agent-instructions}

See Figure \ref{fig:lowkillrate-agent-instructions}.

\begin{figure*}[tb]
\footnotesize
\noindent\fbox{\begin{minipage}{\dimexpr\linewidth-2\fboxsep-2\fboxrule\relax}
Use the following Category vocabulary:

\begin{itemize}
\tightlist
\item
  \textbf{Weak-value/loose-bound oracle}: the changed/relevant code genuinely executes, but the
  test's assertion doesn't distinguish old (buggy) from new (fixed) behavior for the specific
  input the test supplies.
\item
  \textbf{Wrong entry point / bypassed wrapper}: the test's input/setup routes execution around
  the specific code path that changed, so the change is never actually exercised by this test
  (even though a superficially similar path is).
\item
  \textbf{Dead assertion order}: a compound \texttt{\seqsplit{with\ (A,\ B):}} block where
  Python's exit order means one of the two checks can never actually fail.
\item
  \textbf{Operator-scope gap / short-circuit-masked variant}: the surviving mutant sits behind
  Python \texttt{\seqsplit{and}}/\texttt{\seqsplit{or}} short-circuiting, so the mutated sub-expression never actually executes
  even though coverage.py credits the line -- or the mutant is on an AST node type outside the
  mutation-operator vocabulary used here (comparison/arithmetic/boolean only) to begin with.
\item
  \textbf{Commit-scope contamination}: the surviving mutant sits in a
  commit's unrelated files, bundled into the same scored diff.
\item
  \textbf{Import-time pseudo-coverage artifact}.
\item
  \textbf{No defect}: the test is well-written and correctly checks what it claims; a low kill
  rate does NOT reflect any flaw in the test.
\end{itemize}

A known pattern to watch for specifically: many low-kill-rate tests assert a NEGATIVE/boundary
case (e.g.~``returns empty'', ``returns None'', ``is a no-op'', ``rejected''). These are frequently
legitimate -- a mutation applied to a \emph{different} branch/case than the one the test's
negative-boundary input reaches will trivially survive without that being any flaw in the
test. Verify this per-test by actually reading the diff and the mutated code locations, don't
assume either way.

For every survived mutant, read the actual test body AND the actual production code, trace
whether the mutated line is truly on the path the test's assertions depend on, and where
useful, apply the mutation directly (hand-patch a scratch copy, or use the project's real
mutation machinery) and rerun the test to settle ambiguous cases empirically rather than by
inference alone.

What to do for each test:

\begin{enumerate}
\def\labelenumi{\arabic{enumi}.}
\tightlist
\item
  Read the test body at its introducing commit, and the production code it exercises.
\item
  Pull the exact list of mutants for it (operator, file, line, killed/survived) -- don't just
  trust the aggregate number.
\item
  For each SURVIVING mutant: explain the mechanism using the category vocabulary above, or
  propose a new category if genuinely none fits -- name it precisely and explain why it
  doesn't fit the existing ones.
\item
  Give a verdict per test: \textbf{No defect} (survivors are a methodology artifact) or \textbf{Genuine
  defect} (the test's oracle really is weak/loose and would miss a real regression) -- for
  genuine defects, be precise about what the exact gap is and what input would expose it.
\end{enumerate}

Be rigorous -- every claim needs to be independently verified against the actual diff and test code,
not guessed.
\end{minipage}}
\caption{Low-kill-rate defect-mining agent instructions (verbatim text given to the triage agents).}\label{fig:lowkillrate-agent-instructions}
\end{figure*}

\footnotesize

\begin{table}[H]
\footnotesize
\centering
\caption{How Extractor test-writing episodes were prompted, by pattern.}\label{tab:test-writing-patterns}
\begin{tabular}{@{}
  >{\raggedright\arraybackslash}p{(\linewidth - 4\tabcolsep) * \real{0.6461}}
  >{\raggedleft\arraybackslash}p{(\linewidth - 4\tabcolsep) * \real{0.1591}}
  >{\raggedleft\arraybackslash}p{(\linewidth - 4\tabcolsep) * \real{0.1948}}
@{}}
\toprule
Pattern & n & \% \\
\midrule
BUG\_\allowbreak FIX\_\allowbreak REGRESSION & 45 & 37\% \\
OTHER & 33 & 27\% \\
FEATURE\_\allowbreak IMPLEMENTATION\_\allowbreak UNPROMPTED & 21 & 17\% \\
EXPLICIT\_\allowbreak TEST\_\allowbreak REQUEST & 16 & 13\% \\
COVERAGE\_\allowbreak REVIEW\_\allowbreak REQUEST & 6 & 5\% \\
\bottomrule
\end{tabular}
\normalsize
\end{table}

\normalsize

\subsection{How was Extractor test writing prompted?}\label{sec:prompting}

Every Claude Code session transcript from the Extractor code development history was mined for episodes
where a test file was created or modified. The set of episodes was then analysed to extract common patterns. The following patterns, summarised in Table \ref{tab:test-writing-patterns}, are observed:

\emph{BUG\_FIX\_REGRESSION}. A bug was found and fixed, and a regression test was added to pin the
fix without a separate request for tests. Bugs were originally surfaced by: (i)
the human instruction names a TODO number or points at an
existing notes file the coding agent itself maintains e.g.~``go ahead with \#44'', (ii) a short, fresh question or observation from the human user not tied to anything
pre-logged, e.g.~``why does it stall/fail?'', (iii) the agent found the bug itself mid-task.

\emph{OTHER}. Splits roughly evenly between: (i) a refactor's own stated constraint
(``don't break functionality'') led the agent to write tests, and (ii) tests were modified
because a file moved, a function was deleted, or a public API changed shape.

\emph{FEATURE\_IMPLEMENTATION\_UNPROMPTED}. A new feature or function was requested with no
mention of tests anywhere in the instruction, and the agent wrote tests while delivering the work.

\emph{EXPLICIT\_TEST\_REQUEST}. The user directly asked to write tests, add regression coverage, or
verify something works e.g.~``yes, go ahead with that plan. add tests to ensure good code coverage as you go along.'', ``as usual add tests if needed to avoid regression later''

\emph{COVERAGE\_REVIEW\_REQUEST}. The user asked specifically to review test coverage and improve
it, as distinct from asking for a test alongside one feature e.g.~``check the test coverage of the new code, add extra tests if needed.''

Only 22\% of episodes (EXPLICIT\_TEST\_REQUEST + COVERAGE\_REVIEW\_REQUEST) trace to a direct ask for
tests; the remaining 78\% were the agent's own initiative.

\end{document}